\documentclass[3p]{elsarticle}
\usepackage[latin1]{inputenc}
\usepackage{bbm}
\usepackage{graphicx}
\usepackage{cancel}
\usepackage{lmodern}
\usepackage{color}
\usepackage{bbm}
\usepackage{lipsum}
\usepackage{amsmath}
\usepackage{amsfonts}
\usepackage{amssymb}
\usepackage{mathrsfs}
\newcommand\beq{\begin{equation}}
\newcommand\eeq{\end{equation}}
\newcommand\bear{\begin{eqnarray}}
\newcommand\eear{\end{eqnarray}}
\usepackage{epstopdf}
\usepackage{pdfpages}
\usepackage{setspace}
\usepackage{color}

\usepackage{tabularx,ragged2e,booktabs,caption}
\journal{JMPS}
\begin{document}
\begin{frontmatter}
\title{ 
A comprehensive lattice-stability limit surface for graphene}
\author{Sandeep Kumar }
\ead{lahirisd@mit.edu}
\author{David M. Parks}
\ead{dmparks@mit.edu}
\address{Department of Mechanical Engineering \\ Massachusetts Institute of Technology, Cambridge, MA 02139}
\begin{abstract} 
 The limits of reversible deformation in graphene under various loadings are examined using lattice-dynamical stability analysis. This information is then used to construct a comprehensive lattice-stability limit surface for graphene, which provides an analytical description of incipient lattice instabilities of \textit{all kinds}, for arbitrary deformations, parametrized in terms of symmetry-invariants of strain/stress. Symmetry-invariants allow obtaining an accurate parametrization with a minimal number of coefficients. Based on this limit surface, we deduce a general continuum criterion for the onset of all kinds of lattice-stabilities in graphene: an instability appears when the magnitude of the deviatoric strain $\gamma$ reaches a critical value $\gamma^c$ which depends upon the mean hydrostatic strain $\bar {\mathcal E}$ and the directionality $\theta$ of the principal deviatoric stretch with respect to reference lattice orientation. We also distinguish between the distinct regions of the limit surface that correspond to fundamentally-different mechanisms of lattice instabilities in graphene, such as structural versus material instabilities, and long-wave (elastic) versus short-wave instabilities. Utility of this limit surface is demonstrated in assessment of incipient failures in defect-free graphene via its implementation in a continuum Finite Elements Analysis (FEA). The resulting scheme enables on-the-fly assessments of not only the macroscopic conditions (e.g., load; deflection) but also the microscopic conditions (e.g., local stress/strain; spatial location, temporal proximity, and nature of incipient lattice instability) at which an instability occurs in a defect-free graphene sheet subjected to an arbitrary loading condition. 
\end{abstract}
\begin{keyword} 
Graphene, ideal strength, lattice-stability limits, finite element analysis.
\end{keyword}
\end{frontmatter}
\section{Introduction}
Limits to reversible deformation---the stress or strain at which elastic-to-inelastic transition takes place in a material --- pose  fundamental constraints on a material's performance and determine its strength. The limiting stress or strain, in general, depends upon the loading path, and the dependence is often described by means of a phenomenological model termed a limiting (or failure) criterion. Mathematically, a limiting criterion is represented as a surface in stress or strain space, which separates the stable states of reversible deformation  from  the `failed' or irreversibly-deformed states.  For many  materials, the limiting criterion is simple enough to be characterized by one or two material constants, which are readily determined from experiments. Examples include the Mises (or Tresca) yield criterion for metals, the Mohr-Coulomb criterion for cohesive-frictional solids \cite{coulomb1776essai}, the Drucker-Prager criterion for pressure-dependent solids \cite{drucker1952soil}, and the Hoek-Brown criterion for rocks \cite{Hoek1997practical}.

In crystalline materials that are free from lattice imperfections, the limit to elastic deformation sets an upper bound on the material strength \cite{macmillan1983the}. This upper bound, termed the ideal strength, depends on the intrinsic nature of bonding between atoms in the material.   Because of the various types of lattice defects, such as dislocations, grain boundaries, interstitial impurities, and voids which normally exist in materials \cite{minor2006new, suresh2008deformation}, most conventional materials are irreversibly-deformed at stress-levels well below the ideal strength. However, in recent years, advancements in nanotechnology have  enabled fabrication and growth of defect-free two-dimensional crystals in which mechanical failure indeed occurs upon reaching stress levels near the ideal strength \cite{chuang2009near, wang2013near}.

 Graphene, an atomic monolayer comprising a hexagonal network of covalently-bonded C-atoms, is a representative example of such materials (see Fig.~(\ref{BZ})). Experimental studies have shown that defect-free, single-crystalline graphene can sustain near-ideal-strength stresses while remaining within the reversible regime of deforequationmation \cite{lee2008measurement, rasool2013measurement}. Beyond the limit of elastic deformation, the fate of the material is determined by a strength-limiting mechanism such as incipient plasticity or crack-initiation.  Since, in the absence of lattice-imperfections, a strength-limiting mechanism can only be activated by a lattice-instability \cite{li2002incipient, liu2007ab}, the incipient failure of a defect-free crystalline material is intrinsically related to the loss of internal lattice-stability. The point at which the loss of lattice-stability occurs is called the \textit{lattice-stability limit}, and it varies from loading path to path. Further, the dependence of the lattice-stability limit on the loading condition in crystalline materials is generally too complex to be adequately characterized by one or two parameters.
 
Individually, lattice-instabilities of all kinds can be assessed via the lattice-dynamical stability analysis (see Born \& Huang \cite{born1954dynamical}; Hill \cite{hill1962acceleration}), which asserts that a necessary and sufficient criterion for an ideal crystal under arbitrary uniform loading to be stable is that it exhibits stability with respect to bounded perturbations of all wavelengths. Integration of the lattice-dynamical stability analysis with a continuum analysis scheme such as FEA would be ideal for failure analysis of defect-free graphene crystals. This could enable assessment of  incipient failures and ideal strength of graphene, under arbitrary loading conditions at realistic length-scales and slow enough loading rates, directly in a continuum-level simulation. However, stability analysis  based on lattice-dynamics requires carrying out an elaborate sequence of computationally-expensive steps that can not be treated within the confines of an analytical framework, making it difficult to integrate the lattice-dynamical stability analysis into a continuum scheme. Therefore, there remains a need for a general continuum criterion that could describe the onset of instability (of any kind) in graphene under an arbitrary state of deformation.

The aim of this work is to construct a comprehensive lattice-stability limit surface, which constitutes an analytical parametrization of incipient lattice instabilities of \textit{all kinds}, over the space of all homogeneous deformations, in terms of stress/strain. There are two main difficulties in obtaining such a parametrization: First,  crystalline materials are intrinsically anisotropic, so material response, including lattice-stability limit, varies with orientation \cite{Neumann}. Secondly, two fundamentally-different types of lattice-instabilities  govern strength-limiting mechanisms under different loading conditions \cite{marianetti2010failure, clatterbuck2003phonon}: a long-wave (or elastic) instability and a short-wave (or soft mode) instability. The condition for onset of an elastic instability can be parameterized in terms of strain via acoustic tensor analysis (see Kumar \& Parks \cite{kumar2014on} for details), whereas  the short-wave instabilities are much more complex since there is no continuum framework for parametrization of limiting conditions governing the onset of short-wave instabilities. 

The proposed parametrization, in order to overcome the above-mentioned difficulties, employs interpolation of the lattice-stability limits of graphene, corresponding to some representative homogeneous deformation modes, in the basis of symmetry-invariants of the strain/stress measure. Symmetry-invariants are special scalar functions of strain/stress that remain invariant under the point group operations. The use of symmetry-invariants ensures that the parametrization possesses the appropriate material anisotropy of graphene. It also introduces substantial functional simplification, reducing the requisite representative deformation modes needed for interpolation to a small set of biaxial deformations along  the two special symmetry-directions in graphene: armchair and  zigzag. The individual lattice-stability limits used in the interpolation are obtained by atomistic-level lattice-dynamical stability analysis based on phonons to ensure that lattice instabilities of all kinds are captured in the analysis. 

This paper is organized as follows. In Sec.(\ref{schemeofthings}), we briefly summarize the kinematics of graphene, outline the general framework of  invariant-based representation theory of anisotropic materials,  explicitly derive  symmetry-invariants of the strain measure  with respect to the symmetry-group of graphene, and discuss the implementation of these ideas in the context of an (incipient) instability function of graphene. Employing a representation of the ideas outlined  in Sec.(\ref{schemeofthings}), we systematically determine an analytical form for the limit surface for graphene in Sec.(\ref{specializationtographene}). Then, in Sec.(\ref{straintostressspace}), we map the instability function from strain space to stress space.  The model is validated in Sec.(\ref{validation}): in Sec.(\ref{numericalaccuracy}), we assess the numerical accuracy of the representation of failure model based on interpolation; in Sec.(\ref{acoustictensor}), we compare the material instabilities predicted by the failure model with those from acoustic tensor analysis; and in Sec.(\ref{buckling}), we compare the model predictions for buckling instabilities against the results obtained from buckling stability analysis based on the hyperelastic  constitutive relation.  FEA implementation of the failure function is discussed in Sec.(\ref{FEA}): we discuss the procedure in Sec.(\ref{outline}), and illustrate its use in an example of limiting blister-type deformation in Sec.(\ref{example}).  Finally, we conclude in Sec.(\ref{conc}) by summarizing our results and their implications for analysis of strength-measuring nano-scale contact experiments.

\section{Framework of representation}
\label{schemeofthings}
\subsection{Kinematics}
We consider graphene as a 2D deformable body denoted by unstressed reference configuration $\mathcal B$. Let $\mathbf X$ and $\mathbf x$ denote the coordinates of a material element of graphene in undeformed and deformed configuration, respectively.  The convection of material points under deformation is described by a smooth, injective (one-to-one) function $\chi(\mathbf X, t)$ called the motion. The non-translational part of the motion can be equivalently defined by the positive-definite  second-order deformation gradient tensor, $\mathbf F = \nabla \chi(\mathbf X, t) $. Then, the polar decomposition theorem provides the following factorizations of $\mathbf F$ \cite{gurtin2010mechanics, gurtin1982introduction, fung1977first}:
\begin{equation}
\mathbf F= \mathbf R \mathbf U =   \mathbf V\mathbf R,
\end{equation}
where the orthogonal tensor $\mathbf R \in SO_2$ characterizes  rigid-body rotation, and $\mathbf U$ (or $\mathbf V= \mathbf{RUR}^{T}$), termed the right (left) Cauchy-Green tensor, characterizes shape- and area-change. The deformation of a material point can be kinematically factored  as the product of a purely dilatational (or shape-preserving, but area-changing) deformation $\mathbf U^{a}$, and a purely isochoric (or shape-changing, but area-preserving) deformation $\tilde{\mathbf U}$. Accordingly, the stretch tensor can be product-decomposed as
\begin{equation}
      \mathbf U = \mathbf{U}^{a} \tilde{\mathbf U} =  \tilde{\mathbf U} \mathbf{U}^{a},
\end{equation}
where
\begin{equation}
   \mathbf{U}^{a}  \equiv J^{1/2} \, \mathbf{I}
\end{equation}
and 
\begin{equation}
   \tilde{\mathbf U} \equiv \lambda \mathbf r_1 \otimes \mathbf r_1 +\lambda^{-1}\mathbf r_2 \otimes \mathbf r_2;
\end{equation}
here $J = \mathrm{det}\, \mathbf U,\ \lambda \ge 1$ is the deviatoric stretch, $\mathbf I$ is the 2D identity tensor, and $\mathbf r_1= \cos \theta\ \mathbf e_1 +  \sin \theta\ \mathbf e_2$ and $\mathbf r_2= -\sin \theta\ \mathbf e_1 +  \cos \theta\ \mathbf e_2$ are the major and minor principal stretch directions, respectively.\\ \\
Graphene is a composite of two Bravais lattices shifted with respect to each other by a shift vector $\mathbf d_0$. Upon deformation, the individual Bravais lattices follow the macroscopic deformation gradient $\mathbf F$, i.e., the lattice vectors in the deformed crystal are given as  $\mathbf a_1' = \mathbf F \mathbf a_1$ and $\mathbf a_2'=\mathbf F \mathbf a_2$. This is called Cauchy-Born kinematics. However, the shift vector, separating the two lattices, does not obey the Cauchy-Born kinematics, i.e., $\mathbf d \ne \mathbf F \mathbf d_0$. The shift vector $\mathbf d$, in a deformed state, is determined by minimization of total energy of the deformed crystal. That is, under certain imposed deformations, graphene experiences sub-lattice shifts that lower the total energy of the crystal compared to the Cauchy-Born unrelaxed crystal. These deformation modes are the ones that involve a deviatoric stretch, i.e., $\tilde{\mathbf U} \ne \mathbf I$. Examples of this class of deformations include the uniaxial stress/stretch and simple shear. Conversely, because of symmetry, a purely volumetric deformation does not generate a sub-lattice shift.\\
The sub-lattice shift $\mathbf s$, associated with a deformation gradient $\mathbf F$ is defined as the difference between the $\mathbf d$ vectors connecting the two Bravais lattices in the relaxed and unrelaxed configurations. Mathematically,
\begin{equation}
\mathbf s = \mathbf d - \mathbf d_{CB}; \text{where } \mathbf d_{CB} = \mathbf F \mathbf d_0, 
\end{equation}
where $\mathbf d_0$ denotes the shift-vector in the undeformed configuration.\\
The rigorous energetic and kinematic continuum description of a complex crystal, such as graphene, requires the definition of the strain energy density function as follows
\begin{equation}
\psi = \check{\psi}(\mathbf F, \mathbf s).
\end{equation}
The above description implies the following variational form
\begin{equation}
\delta \psi = \frac{\partial \psi}{\partial \mathbf F} : \delta \mathbf F + \frac{\partial \psi}{\partial \mathbf s} : \delta \mathbf s.
\end{equation}
Localization of the principle of virtual work then requires the following energy balance
\begin{equation}
\mathbf 0 = \bigg \lgroup \frac{\partial \psi}{\partial \mathbf F} - \mathbf T^{(1)} \bigg \rgroup : \delta \mathbf F + \bigg \lgroup \frac{\partial \psi}{\partial \mathbf s} - \mathbf f \bigg \rgroup: \delta \mathbf s,
\label{w-c-localization}
\end{equation}
where $\mathbf T^{(1)}$ is the first Piola-Kirchhoff stress (stress tensor work-conjugate to $\mathbf F$), and $\mathbf f$ are the forces per unit reference volume acting on the atoms of the state defined by ($\mathbf F$, $\mathbf s$). The conditions (\ref{w-c-localization}) imply the equilibrium equations
\begin{equation}
\frac{\partial \psi}{\partial \mathbf F} = \mathbf T^{(1)},
\end{equation}
and
\begin{equation}
\frac{\partial \psi}{\partial \mathbf s} = \mathbf f.
\end{equation}
At equilibrium, the force on each atom in the lattice is zero, and the above equation becomes
\begin{equation}
\frac{\partial \psi}{\partial \mathbf s} = \mathbf 0.
\label{shiftvector}
\end{equation}
Thus equation~(\ref{shiftvector}) implicitly determines the equilibrium value of sub-lattice shift, $\mathbf s = \mathbf s_{\text{eq}} (\mathbf F)$, for which the crystal satisfies both external and internal equilibria. Inserting this form into the strain energy function allows representation of the equilibrium strain energy in terms of
$\mathbf F$ alone:
\begin{equation}
\psi = \check{\psi}(\mathbf F, \mathbf s_{\text{eq}}(\mathbf F)) \equiv \hat{\psi}(\mathbf F).
\label{F_only}
\end{equation}
\subsection{Strain measure}
In this work, the logarithmic strain measure $\mathbf E^{(0)} = \ln \mathbf U$ is employed in the representation of the stress-strain constitutive response as well as of the limit surface. The spectral representation of  $\mathbf E^{(0)} \equiv \ln \mathbf U = \ln \mathbf U^{a} + \ln \tilde{\mathbf U}$  is given by:
\begin{eqnarray}
\mathbf E^{(0)} &=& \underbrace{\frac{1}{2}\,  \ln J \, \mathbf I}_{  \ln  \mathbf U^{a}} + \underbrace{\ln \lambda \, \left(\mathbf r_1 \otimes \mathbf r_1 - \mathbf r_2 \otimes \mathbf r_2 \right)}_{  \ln \, \tilde{\mathbf U}}  \nonumber \\ 
                    & \equiv &  \overbrace{\, \frac{1}{2} \, \, \epsilon_a \, \mathbf{I}}  \quad +   \overbrace{\quad \quad  \quad \quad \mathbf{E}^{(0)}_{0} ,\quad \quad \quad}
 \label{log}
\end{eqnarray}
where  
\begin{equation}
\epsilon_a = \text{tr}\, \mathbf E^{(0)} = \ln J = \ln (\det\, \mathbf U),
\end{equation}
gives the areal logarithmic strain $\epsilon_a$, and 
\begin{equation}
\mathbf E^{(0)}_0 = \ln \tilde{\mathbf U} = \ln \lambda \, \left(\mathbf r_1 \otimes \mathbf r_1 - \mathbf r_2 \otimes \mathbf r_2 \right),
\label{sprep}
\end{equation}
denotes the deviatoric part of  $\mathbf {E}^{(0)}$. The orientation $\theta$ is measured from a fixed material axis aligned with the zigzag direction of the graphene lattice. In the rest of this article, we will refer to $\theta$ as the \textit{deviatoric stretch angle}.
Alternatively, we may also write the spectral representation of the logarithmic strain as
\begin{equation}
\mathbf E^{(0)} =\mathcal E_{\text{max}}\, \mathbf r_1\otimes \mathbf r_1 + \mathcal E_{\text{min}}\, \mathbf r_2 \otimes \mathbf r_2.
\end{equation}
where $\mathcal E_{\text{max}} = \ln(J^{1/2} \lambda)$ and $\mathcal E_{\text{min}} =\ln(J^{1/2}/ \lambda) $ are the major and minor principal logarithimic strains, respectively.
 In addition, we also define a mean hydrostatic strain $\bar{\mathcal E} = \frac{1}{2} (\mathcal E_{\text{max}} + \mathcal E_{\text{min}}) = \ln J^{1/2}$, which characterizes the areal dilatation or contraction of the material.
\subsection{Symmetry-constraints on the limit surface}
The proposed approach is based on the notion of a limit function --- a continuous scalar-valued function $\mathcal F$ of the strain measure $\mathbf E^{(0)}$ --- such that all  deformed configurations of the lattice that lie on the surface
\begin{equation}
\mathcal F(\mathbf E^{(0)} )=0,
\end{equation}
have reached a state of incipient lattice instability.  All deformed states of the crystal  lying in the interior of the limit surface satisfy $\mathcal F(\mathbf E^{(0)}) > 0$, and are stable with respect to lattice perturbation of all wavelengths; conversely, for deformed states lying outside the surface, i.e., $\mathcal F(\mathbf E^{(0)}) < 0$, the lattice is unstable with respect to an incremental perturbation of some wavelength.\\ 
For an anisotropic material, the limit function $\mathcal F$ --- in accordance with Neumann's principle \cite{Neumann} --- should include the material symmetry group $\mathcal G$ of the underlying lattice, i.e., 
\begin{equation}
\mathcal F(\mathbf Q^T \mathbf E^{(0)} \mathbf Q) = \mathcal F(\mathbf E^{(0)} )\ \forall \ \mathbf Q \in \ \mathcal G,
\label{isotropicization}
\end{equation}
where $\mathbf Q$ is an orthogonal tensor denoting the symmetry operations included in the material symmetry group $\mathcal G$ ($\mathcal C_{6v}$ in the case of graphene). A scalar-valued function of a tensor agency--- such as $\mathcal F(\mathbf E^{(0)})$ --- that remains invariant under a  material symmetry group $\mathcal G$ is called a $\mathcal G$-invariant scalar function. The representation of a generic $\mathcal G$-invariant scalar function involves using the isotropicization theorem and symmetry-invariants of the tensor agency with respect to the point symmetry group of the material.\\ 
\subsection{Isotropicization theorem and symmetry-invariants}
 The isotropicization theorem --- based on the notion of a materially-embedded structure tensor $\mathbb H$ --- allows  a $\mathcal G$-invariant scalar function to be expressed in terms of a list of special scalar functions --- $\mathcal J_1$, $\mathcal J_2$, ..., $\mathcal J_n$ --- which are joint isotropic functions of $\mathbf E^{(0)}$ and $\mathbb H$ \cite{boehler1979simple, boehler1977irreducible, lokhin1963nonlinear}; i.e., 
\begin{equation}
\mathcal F (\mathbf E^{(0)}) = \hat {\mathcal F} (\mathcal J_1, \mathcal J_2, ..., \mathcal J_n), 
\end{equation}
where
\begin{equation}
       {\mathcal J}_i(\mathbf{E}^{(0)}; \, \mathbb H) =  {\mathcal J}_i(\mathbf{Q}^T\mathbf{E}^{(0)}\mathbf{Q}; \, {\mathbb P}_{\mathbf{Q}}(\mathbb H))
      \, \forall \ \mathbf{Q} \in SO_2.
        \end{equation}
      Here  $ {\mathbb P}_{\mathbf{Q}}$ denotes the  transformation of the structure tensor $\mathbb H$ under the orthogonal transformation $\mathbf{Q}$.
The functions  $\mathcal J_i$  are termed symmetry-invariants since they satisfy all the constraints belonging to the material symmetry group of the crystal. Smith \cite{Smith1, Smith2, Smith3} showed that the set of mutually-independent symmetry-invariants  serves as a  complete and irreducible basis for the representation of scalar constitutive functions of the anisotropic material. In the following section, we explicitly derive symmetry-invariants of $\mathbf E^{(0)}$ for the structure tensor characterizing the material symmetry group of graphene.
\subsection{Symmetry-invariants and functional basis for $\mathcal C_{6v}$ material symmetry group }
 \begin{minipage}{\linewidth}
 \centering
\includegraphics[scale=0.6]{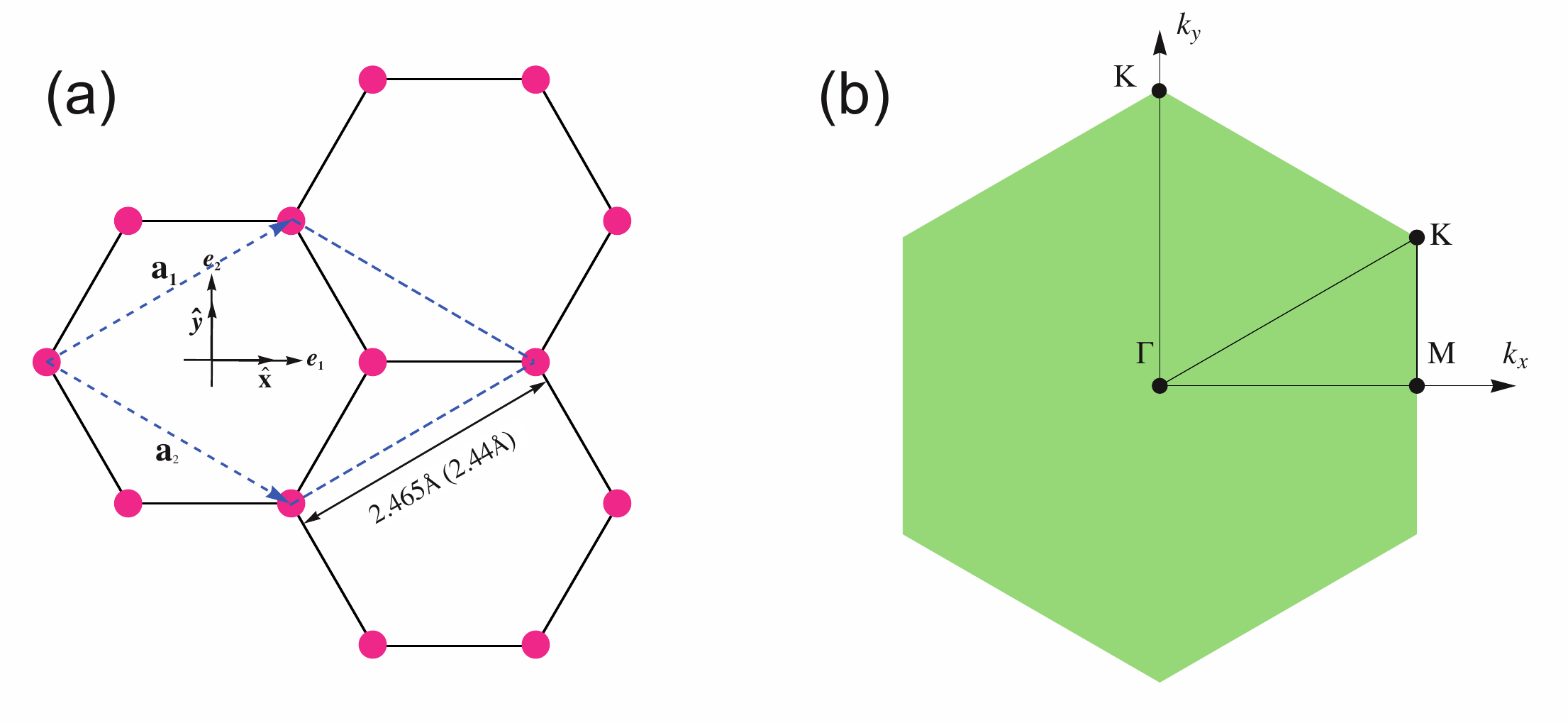}
 \captionof{figure}{\small (a) Graphene lattice with orientations of the material unit vectors --- $\hat{\mathbf x}$ and $\hat{\mathbf y}$ --- and the Cartesian unit vectors --- $\mathbf e_1$ and $\mathbf e_2$ --- indicated. The dashed blue lines denote the unit cell used in the \textit{ab initio} calculations. The GGA (LDA) value of the lattice parameter is also indicated. The armchair and zigzag directions are along the $\mathbf e_1$ and $\mathbf e_2$ axes, respectively. (b) Brillouin zone of graphene with high symmetry points and irreducible wedge indicated. }
 \label{BZ}
 \end{minipage}
 \\ \\
The structure tensor characterizing the $\mathcal C_{6v}$ point group, the material symmetry group of graphene, is a sixth-order tensor given by \cite{zheng1993tensors, zheng1993two}
\begin{equation}
\mathbb H = \mathbf M \otimes \mathbf M \otimes \mathbf M  - \left(\mathbf M \otimes \mathbf N \otimes \mathbf N +\mathbf N \otimes \mathbf M \otimes \mathbf N + \mathbf N \otimes \mathbf N \otimes \mathbf M \right),
\end{equation}
where $\mathbf M$ and $\mathbf N$ are two traceless second order tensors given by
 \begin{equation}
\mathbf M =  \hat{\mathbf x} \otimes\hat{\mathbf x} - \hat{\mathbf y} \otimes\hat{\mathbf y}; \quad \quad 
\mathbf N =  \hat{\mathbf x} \otimes\hat{\mathbf y} + \hat{\mathbf y} \otimes\hat{\mathbf x}.
\end{equation}
$\hat{\mathbf x}$ and $\hat{\mathbf y}$ denote orthogonal material unit vectors fixed in the frame of the  reference crystal such that at least one of them is aligned with an axis of reflection symmetry.\\
The complete and irreducible set of polynomial joint invariants of $\mathbf E^{(0)}$ and $\mathbb H$ constitute what is called an integrity basis, and are given by
 \begin{equation}
 \mathcal J_1 \equiv \epsilon_a= \mathrm{tr} \mathbf E^{(0)}  = \ln J,
 \end{equation}
 \begin{equation}
  \mathcal J_2 \equiv   (\gamma_i/2)^2= \frac{1}{2}\ \mathbf E^{(0)}_0 : \mathbf E^{(0)}_0 =  (\ln \lambda)^2,
  \end{equation}
  where $\mathbf A : \mathbf B = \mathrm{tr}(\mathbf A^{T} \mathbf B)$ is the scalar tensor product,
  and\\
  \begin{align}
  \mathcal J_3 \equiv (\gamma_{\theta}/2)^3 & = \frac{1}{8}\mathbb H[\mathbf E^{(0)}_0,\mathbf E^{(0)}_0,\mathbf E^{(0)}_0]\\
  & = \frac{1}{8}\left[ \left(\mathbf{M} : \mathbf{E}_0^{(0)}\right)^3 -3\left(\mathbf{M} :\mathbf{E}_0^{(0)}\right) 
  \left(\mathbf{N} : \mathbf{E}_0^{(0)}\right)^2 \right] \\
  &= \left(\ln \lambda \right)^3 \, \cos 6 \theta,
  \end{align}
where $\cos \theta = \mathbf r_1 . \hat{\mathbf x}$ indicates the orientation of maximum principal stretch. 
The first two of these invariants, $\epsilon_a$ and $\gamma_i^2$, are simply two isotropic invariants of $\mathbf E^{(0)}$ alone. Thus any material anisotropy in the constitutive response of graphene is captured solely  by the third invariant $\gamma_{\theta}^3$. In a previous work \cite{kumar2014on}, we showed that a constitutive function of graphene --- such as its finite deformation hyperelastic response --- is conveniently represented in terms of the integrity bases of the logarithmic strain with respect to the material symmetry group of graphene.\\
For the purpose of representation of $\mathcal F$, we employ an alternate set of symmetry-invariants of $\mathbf E^{(0)}$, comprising the mean hydrostatic strain $\bar{\mathcal E}$, the magnitude of deviatoric strain $\gamma$; and the symmetry-reduced principal stretch direction defined as $\Theta(\theta)= \arccos(\cos 6\theta)$. Note that the elements of this set are not independent from the elements of the integrity basis, i.e., $\epsilon_a$, $\gamma_i^2$, and $\gamma_{\theta}^3$; and utilizing kinematical definitions of equations~(\ref{log}-\ref{sprep}), the two sets of symmetry-invariants can be shown to be related as follows: 
 \begin{align}
\hspace*{-3.7cm}\text{1. Mean hydrostatic strain }\bar{\mathcal E} &=\frac{ \epsilon_a}{2}  = \ln J^{1/2}.
 \end{align}
 \begin{align}
\text{2. Magnitude of deviatoric strain } \gamma & = \sqrt{\frac{\mathbf E^{(0)}: \mathbf E^{(0)} }{2}}= \frac{\gamma_i}{2} = |\ln \lambda| \ge 0. \end{align}
  \begin{align}
\hspace*{-2.5cm}\text{3. Directionality }\Theta(\theta) & = \arccos\left [ \frac{\gamma_\theta^3}{\gamma_i^{3}}\right]=  \arccos(\cos 6 \theta).
  \end{align}
  Thus, the failure function has the representation
  \begin{equation} \mathcal F(\mathbf E^{(0)}) = \hat{\mathcal F}(\gamma, \bar{\mathcal E}, \Theta(\theta) ). \end{equation}
The set of symmetry-invariants, $\gamma$, $\bar{\mathcal E}$ and $\Theta(\theta)$, which involve non-polynomial combinations of strain components, is called a \textit{functional basis}. \\
\subsection {Features of the limit surface}
The limit surface ---  as expressed in terms of the elements of the functional basis --- prescribes the critical value of the magnitude of deviatoric strain $\gamma^c$ as a function of the areal strain $\bar{\mathcal E}$ and the symmetry-reduced principal stretch direction $\Theta(\theta)$. Thus, the form of the limit function reads as 
\begin{equation}
\mathcal F(\mathbf E^{(0)}) \equiv \gamma^c(\bar{\mathcal E}, \Theta(\theta)) - \gamma; \ \gamma \ge 0,
\label{fs}
\end{equation}
such  that if, at a given state of deformation, $\mathcal F (\mathbf E^{(0)}) >0 $, then the material is stable with respect to  incremental perturbations of all wavelengths; otherwise the instantaneous configuration of the lattice is unstable under some incremental perturbation. The limit surface, when graphically expressed in terms of $\bar{\mathcal E}$, $\gamma$ and $\Theta(\theta)$, constitutes a 3D polar surface with $\bar{\mathcal E}$ on the radial axis, $\gamma$ on the vertical axis and $\theta$ as the angular coordinate.
 \\ \\
\textbf{Boundedness of the limit surface ---}  The proposed limit surface is bounded both on the vertical axis (i.e., $\gamma$-axis) and the radial axis (i.e., $\bar{\mathcal E}$-axis), as explained in the following.\\\\
\textbf{On the vertical axis:}
\begin{itemize}
\item By definition, the magnitude of deviatoric strain is a positive semi-definite entity, i.e., $\gamma \ge 0$ at all values of $\bar{\mathcal{E}}$. Thus, the limit surface is bounded from below by the surface $\gamma =0$.
\item From above, the failure surface is bounded by the critical value of the magnitude of deviatoric strain beyond which the crystal is unstable, i.e., $\gamma\le \gamma^c(\bar{\mathcal E}, \Theta(\theta))$.
\end{itemize}
\textbf{On the radial axis:}
\begin{itemize}
\item 
Because of graphene's limiting thickness and extreme bending compliance, essentially any attempt to decrease the area of a sufficiently large, flat graphene sheet would result in out-of-plane buckling, graphene's small, though non-zero bending stiffness notwithstanding; for this reason, we consider
that the  domain of  the surface is bounded from below: $\bar{\mathcal E} \ge 0$, providing a graphical representation using
$\bar{\mathcal E}$ as a `radial' coordinate.  Moreover,  if 
$\bar{\mathcal E}$  is held at $0$, imposition of a non-zero $\gamma$  creates a negative principal stress, again leading to buckling in a sufficiently large, flat graphene sheet; hence we take $\gamma^c (\bar{\mathcal E}=0, \Theta(\theta))=0$ for all $\theta$.  (These limits follow from the
quadratic phonon dispersion relations of ZA phonons at $\Gamma$ in a periodically-extended planar crystal.)
\item Upon a continued equi-biaxial deformation, a state is reached, $\bar{\mathcal E} =\bar{\mathcal E}_{\text{max}} $ when graphene fails without any shear strain, i.e., in pure dilatation. Therefore, the radial domain of the surface is bounded from above as well: $\bar{\mathcal E} \le \bar{\mathcal E}_{\text{max}}$ and $\gamma^c (\bar{\mathcal E}=\bar{\mathcal E}_{\text{max}}, \Theta(\theta))=0$ for all $\theta$.
\end{itemize}
\section{Form of the limit surface in terms of the functional basis}
\label{specializationtographene}
Employing the representation ideas outlined in the previous section, we now systematically determine the functional form of the limit function $ \mathcal F \equiv  \gamma^c(\bar{\mathcal E}, \Theta(\theta))- \gamma$ for graphene in terms of the three symmetry-invariants.  For this purpose, we first obtain the set of data containing the stability-limiting values $\gamma^c$ over a range of $\bar{\mathcal E}$ values, calculated from lattice-dynamical stability analysis, for a set of representative homogeneous deformation paths.
\subsection{Lattice dynamical stability analysis}
\label{latticedynamics}
The atoms in a crystalline material remain in a state of oscillatory motion about their equilibrium sites, constituting what is called the random lattice vibrations. Phonons are the normal modes of the lattice vibrations. Fundamentally, each phonon is a collective oscillation of atoms characterized by a wave-vector $\mathbf q$ and a oscillation frequency $\omega_n$, where $n$ labels the branch index. The relation between $\omega_n$ and $\mathbf q$ is called the dispersion relation of the crystal.  The phonon-dispersion relation of a crystal contains a total of $3 \mathcal N$ branches where $\mathcal N$ is the number of atoms in the unit cell ($\mathcal N=2$ for graphene). 
Three (3) of these branches are called acoustic and the remaining $3(\mathcal N-1)$ are called optical.\\
Phonons constitute a useful means of assessing the mechanical stability of a crystal lattice. Mechanical stability of a crystal requires that the underlying lattice remains stable against spatial perturbations of all wavelengths, and this is ensured if the following two conditions hold:
\begin{itemize}
\item First, all the acoustic branches should have a positive slope at $\mathbf q=\mathbf 0$, i.e.,
\begin{equation}
\frac{d\omega_n}{d|\mathbf q|} \bigg \rvert_{\mathbf q =\mathbf 0}>0\ \forall \ n  \in 1,..., 3.
\label{acousticfailure}
\end{equation}
If at some point along the loading path, this criterion ceases to hold then it is an indication of an incipient instability, called a long wavelength instability, in the material. In monolayer graphene, a long wavelength instability may correspond to either an out-of-plane instability (buckling) or an elastic instability \cite{liu2007ab, kumar2010intrinsic}. Buckling is a mode of structural instability wherein graphene explores the third spatial dimension via deformation under compression. Unlike a material instability, buckling does not involve material fracturing or undergoing plastic deformation. Further, since graphene is one atomic layer thin, a buckling instability manifests itself as a long-wavelength instability in out-of-plane acoustic phonon (also called ZA branch) mode of graphene.\\ On the other hand, an elastic instability is a mode of material failure and manifests itself as a long-wavelength instability in the in-plane acoustic phonon (LA or TA branch) modes of graphene. An elastic instability in the material can also be captured by acoustic tensor analysis; therefore the onset of elastic instabilities in material can be parameterized with strain via the acoustic tensor route also.
\item Second, all phonon frequencies for $\mathbf q \ne \mathbf 0$ should be real and non-zero, i.e.,
\begin{equation}\omega_n^2(\mathbf q \ne \mathbf 0) > 0\ \forall \ n \in 1,..., 3\mathcal N.
\label{softmodefailure}
\end{equation}
If at some point along the loading path, this criterion ceases to hold then it is an indication of an incipient soft mode instability in the material. Unlike an elastic instability, a soft mode instability does not have an equivalent continuum criterion, and therefore a mathematical framework parametrizing the onset of such instabilities in terms of stress or strain.
\end{itemize}
Lattice-stability analysis based on the above criteria is most effectively assessed by phonons, which, below the Debye temperature ($T_D =1000\,K$ for graphene), constitute a complete normal basis for lattice vibrations. A deformation-induced instability of any kind --- macroscopic or microscopic --- is directly visible in the dispersion of phonons. 
\subsection{Basics of density functional perturbation theory}
 DFPT employs density functional theory (DFT) in conjunction with linear response perturbation theory to calculate the lattice dynamical properties of a crystal (see references \cite{baroni2001phonons, DFPT2, mounet2005first}). In density functional theory (DFT) \cite{hohenberg1964inhomogeneous, kohn1965self}, the ground state energy of a system of interacting electrons moving under the influence of an external potential is obtained by minimization of a universal functional, $\mathcal F[n(\mathbf r)]$, of electronic density $\mathbf n(\mathbf r)$. The electronic density $n_0(\mathbf r)$ that minimizes the energy functional also happens to be the true ground state electronic density of the system.\\
Kohn and Sham \cite{kohn1965self} showed that the original system of interacting electrons can be replaced by an auxiliary system of non-interacting electrons moving under the influence of an effective potential such that the auxiliary system at the ground state possesses the same electronic density as the original system. The motion of  electrons in such an auxiliary system are described by a set of one-electron equations (also called Kohn-Sham equations):
\begin{equation}
\left[-\frac{\hbar^2}{2m} \nabla^2 + V_{\text{Eff}}(\mathbf r) \right] \psi_n(\mathbf r) = \epsilon_n \psi_n(\mathbf r) ,
\label{eq1}
\end{equation}
where $\psi_n$ is a Kohn-Sham orbital and $\epsilon_n$ is the corresponding eigenvalue.
The electronic density is calculated from the Kohn-Sham orbitals using the relation:
 \begin{equation} n(\mathbf r) = \sum_n |\psi_n(\mathbf r)|^2 g(\epsilon_F- \epsilon_i), 
 \label{eq2} \end{equation}
 where $g(\epsilon_F- \epsilon_i)$ is the occupation function, and $\epsilon_F$ is the Fermi energy.
The effective potential  $V_{\text{Eff}}(\mathbf r)$ is a functional of electronic density, and is given by
\begin{equation}
V_{\text{Eff}} (\mathbf r) = V_{\text{IE}} (\mathbf r) +  e^2 \int \frac{n(\mathbf r')}{|\mathbf r - \mathbf r'|} d\mathbf r' + v_{xc} (\mathbf r),
\label{eq3}
\end{equation}
where the first term on the right, $V_{\text{IE}} (\mathbf r)$, is the potential due to interaction between electrons and ionic cores, the second term is the potential due to Coloumb interaction between electrons, and the last term, $v_{xc}(\mathbf r) = \delta \mathcal E_{xc}/\delta n(\mathbf r)$, is the functional derivative of the exchange-correlation energy functional $\mathcal E_{xc}$. Once an explicit form for $\mathcal E_{xc}$ has been determined, the set of equations (\ref{eq1} - \ref{eq3}) can be solved self-consistently to determine the ground state energy and electronic density of the system.\\
Calculation of phonons requires knowledge of interatomic force constants, defined as the derivatives of the ground state total energy of the system $E$ with respect to ionic coordinates $\mathbf R = \{\mathbf R_1, \mathbf R_2, ..., \mathbf R_N \}$. Using Feynman-Hellman theorem, we evaluate such derivatives as:
\begin{equation}
\frac{\partial^2 E(\mathbf R)}{ \partial \mathbf R_I \partial \mathbf R_J} = \int  \left[\frac{\partial n (\mathbf r)}{\partial \mathbf R_J} \right]_{n_0(\mathbf r)}\frac{\partial V_\text{IE} (\mathbf r)}{\partial \mathbf R_I} d\mathbf r + \int n_0(\mathbf r) \frac{\partial^2 V_{\text{IE}}(\mathbf r)}{\partial \mathbf R_I \partial \mathbf R_J} d \mathbf r + \frac{\partial^2V_\text{II} (\mathbf R)}{\partial \mathbf R_I \partial \mathbf R_J},
\end{equation}
where $V_\text{II} $ is the energy corresponding to Coloumbic interaction between ionic cores.
Thus, the evaluation of the interatomic force constants requires ground state electron density $n_0(\mathbf r)$ as well as the first-order correction $\left[\partial n (\mathbf r)/\partial \mathbf R_J \right]_{n_0(\mathbf r)}$. \\
The DFPT method calculates the first-order correction in electronic density from lattice's response to a set of monochromatic lattice perturbations, each characterized by a $\mathbf q$-vector. Provided the amplitude of such perturbations are small enough, the application of perturbation theory allows to recover a set of self-consistent relations for the first-order corrections in electronic density and wavefunctions. Let $\psi_v^{\mathbf k}$ and  $\epsilon_v^k$ be the unperturbed wave function and eigenenergy of an electron with wave-vector $\mathbf k$ and band-index $v$, respectively. First-order correction in the wave function then satisfies the following equation:
\begin{equation}
\left[ -\frac{\hbar^2}{2m} \nabla^2 + V_{\text{Eff}} \right] \Delta \psi_v^{\mathbf k + \mathbf q} = \epsilon_v^k \Delta \psi_v^{\mathbf k + \mathbf q}  - \Lambda^{\mathbf k + \mathbf q} \Delta V_{\text{Eff}} (\mathbf r) \psi_v^{\mathbf k},
\label{eq4}
\end{equation}
where $\Lambda^{\mathbf k + \mathbf q} $ is the projector operator over the manifold of states of wave-vector $\mathbf k + \mathbf q$, and $\Delta V_{\text{Eff}} (\mathbf r)$ is the associated perturbation in the effective potential due to perturbation, given by:
\begin{equation}
\Delta V_{\text{Eff}}(\mathbf r) = \int \frac{\Delta n(\mathbf r')}{|\mathbf r- \mathbf r'|} d^3 \mathbf r' + \Delta n (\mathbf r)  \left[\frac{dv_{xc} (\mathbf r)}{dn} \right]_{n_0(\mathbf r)} + \Delta V_{\text{IE}}(\mathbf r).
\label{eq5}
\end{equation}
Perturbation in the electronic density is obtained as:
\begin{equation}
\Delta n = \frac{4}{V} \sum_k \psi_v^\mathbf k  e^{-i \mathbf q.\mathbf r} \Lambda^{\mathbf k + \mathbf q} \Delta \psi_v^{\mathbf k+ \mathbf q},
\label{eq6}
\end{equation}
$V$ being the volume of the system.
Upon  solving the set of equations (\ref{eq4}-\ref{eq6}) self-consistently, we obtain the first-order corrections in wave function and electronic density of the system. Knowledge of such corrections allows direct access to the dynamical matrix of the system, defined as the Fourier transform of the interatomic force constant matrix, i.e.,
\begin{equation}
\mathbb D_{Ia Jb }(\mathbf q) = \frac{1}{\sqrt{M_I M_J}}\sum_{\mathbf R} \mathbb C_{IaJb} (\mathbf R) e^{-i \mathbf q. \mathbf R};\ \mathbb C_{IaJb}=\frac{\partial^2 E(\mathbf R)}{ \partial \mathbf R_{Ia} \partial \mathbf R_{Jb}},
\end{equation}
where $M_{I(J)}$ denotes the mass of $I(J)^\text{th}$ atom. Phonon frequency $\omega(\mathbf q)$ and the associated eigenmode $\mathbf u_\mathbf{q}$ are then determined by solving the eigenvalue equation:
\begin{equation}
\omega^2(\mathbf q) \mathbf u_\mathbf{q} = \mathbb D (\mathbf q) \mathbf u_\mathbf{q}.
\end{equation} 
\subsection{Details of DFPT calculations}
The exchange correlation energy of electrons is treated with the Local Density Approximation (LDA) of Perdew and Wang (\cite{perdew1992accurate}). The interaction between ionic cores and valence electrons is represented by an ultrasoft pseudopotential \cite{vanderbilt1990soft}. Kohn-Sham wave functions are represented using a plane-wave basis with an energy cutoff of 30 Ry and a charge density cutoff of 300 Ry. Integration over the irreducible Brillouin zone (IBZ) is performed with a uniform $30 \times 30 \times 1$ mesh of $k$-points, and occupation numbers are smeared using the Marzari-Vanderbilt cold smearing scheme \cite{mv} with broadening of 0.03 Ry. Errors in the Cauchy stresses and total energy due to basis-set size, smearing parameter, and $k$-points are converged to less than 0.034 N/m and 0.01 Ry, respectively.\\
The phonon dispersion relations --- of the undeformed and deformed graphene --- are computed via linear response calculations as implemented in the density functional perturbation theory (DFPT) \cite{baroni2001phonons, DFPT2}.  
The dynamical matrix  is calculated on an $8 \times 8 \times 1 $ uniform grid of  $\mathbf q$-points in the IBZ  --- which is then fast-Fourier-transformed to calculate the interatomic force constants (IFC), corrected by the acoustic sum rule to ensure that $\omega(\mathbf q=\mathbf 0)=0$ for all the acoustic branches. The IFC's are then used to interpolate  phonon frequencies over a dense set of $\mathbf q$-points in the IBZ. Both energies and phonon dispersion relations in this work are performed on a two-atom primitive unit cell of graphene shown in  Fig.~(\ref{BZ}). All calculations are fully relaxed in terms of shift vector to ensure conditions of zero atomic force.
The  LA and TA phonon branches, in the neighborhood of $\mathbf q =\mathbf 0$, obey linear dispersion:  $\omega_{LA} = c_{LA} q$ and $\omega_{TA} =c_{TA} q$; where $q = |\mathbf q|$ and $c_{LA} $ and $c_{TA} $ are longitudinal and transverse acoustic wave-velocities, respectively.
As a simple verification of the DFPT phonon calculations, Tab.~(\ref{verification}) compares  longitudinal and transverse acoustic wave-velocities obtained from the phonon dispersion with experimentally-measured values; also, since the G band in the Raman spectra of graphene is associated with the doubly-degenerate 
LO - TO vibrational modes at $\Gamma$, we also compare calculated LO/TO frequencies at $\Gamma$  with the G-band frequency measured in Raman spectroscopy.  \\ \\
\begin{minipage}{\linewidth}
\centering
\begin{tabular}{c c c c c}
\toprule[1.5pt] 
 & $c_{LA} $(km/s)  & $c_{TA} $(km/s)& Raman $G$-peak   \\
\hline 
\hline
\\
Measured value \cite{wirtz2004phonon} & $ 21.3 $ &  $ 13.6$ &  1582 $\text{cm}^{-1}$ \\  \\
\hline
\\
Continuum model (LDA) \cite{kumar2014on} & 21.80  & 13.78  & ---\\ 
 & $\overbrace{\sqrt{E_{2D}/\rho_{2D}}}$ &$\overbrace{\sqrt{\mu_{2D}/\rho_{2D}}}$
\\
%
%
%
\hline
\\
DFPT Phonons in this work (LDA) & 21.15 & 13.50 & 1540 $\text{cm}^{-1}$ & \\ 
& $\overbrace{\frac{d\omega_{LA}}{dq} \bigg \rvert_{q=0}}$ &$\overbrace{\frac{d\omega_{TA}}{dq} \bigg \rvert_{q=0}}$ & \\ \\
\bottomrule[1.65pt]
\end{tabular}
\captionof{table}{In-plane longitudinal and transverse wave velocities calculated from phonon dispersion closely agree with measured values as well as with values calculated from the continuum model  \cite{kumar2014on}. The LO/TO frequencies at $\Gamma$ are also in good agreement with the G-peak of the Raman spectra of graphene. Note: $\rho_{2D}$ is the mass per reference area of graphene which is calculated from the atomic mass of carbon and the undeformed lattice constant of graphene; $E_{2D}$ denotes the elastic modulus in uniaxial strain; and $\mu_{2D}$ denotes the shear modulus. }
\label{verification}
\end{minipage} \\ \\
DFPT allows calculation of phonon dispersion relations with high accuracy at any arbitrarily-deformed state of the graphene lattice, whereas  the errors in phonon frequencies based on empirical potentials can be forbiddingly large, rendering the predictions somewhat unreliable. For example, the sound velocities from an empirical many-body potential in the unstrained state of graphene differed from measured values  \cite{wirtz2004phonon} by nearly 50\% \cite{sergey2012ultimate}; in contrast, the longitudinal and transverse elastic wave-velocities obtained from DFPT phonon calculations differ by less than 1\% from measured values, as shown in Tab.~(\ref{verification}).
\subsection {Sampling scheme} 
\label{samplingscheme}
For the purpose of sampling  deformed states of the lattice, we use the deviatoric stretch angle $\theta$ to designate what we term  a `deformation path'; i.e., increasing $\gamma$ at a fixed $\theta$ is referred to as moving along the deformation path prescribed by $\theta$.
To determine the point of incipient lattice instability along  a deformation path:
\begin{itemize}
\item The lattice is first deformed via a pure equi-biaxial strain, i.e., $\mathbf E^{(0)}_v = \bar{\mathcal E} \mathbf I$ , followed by an isochoric shape-changing stretch of the form $\mathbf E^{(0)}_0= \gamma (\mathbf r_1 \otimes \mathbf r_1 - \mathbf r_2 \otimes \mathbf r_2)$ where $\mathbf r_1 = \cos\, \theta\, \mathbf e_1 + \sin \, \theta\, \mathbf e_2$ and $\mathbf r_2 \perp \mathbf r_1$; $\theta$ is kept fixed. Thus, total strain in the final configuration is $\mathbf E^{(0)} =\mathbf E^{(0)}_v  + \mathbf E^{(0)}_0 $, as schematically shown in Fig.~(\ref{sampling}-a). 
 \item For each deformation pair $\{\bar{\mathcal E},\, \gamma\}$ along the sampled deformation path $\theta$, we check the phonons to determine if the conditions for acoustic stability (given by equation (\ref{acousticfailure})) and short-wave stability (given by equation (\ref{softmodefailure})) hold. If at some critical magnitude of deviatoric strain, $\gamma = \gamma^c$, either of the two conditions ceases to hold then the lattice is considered as incipiently unstable, and the corresponding triplet $\{\bar{\mathcal E},\, \gamma^c,\, \theta\}$ corresponds to a point on the instability surface.  
 \end{itemize}
 Our sampling set includes  values of $\bar{\mathcal E}$ ranging from the undeformed state $\bar{\mathcal E}=0$ to the critical equi-biaxial deformation, $\bar{\mathcal E}_{c} = 0.144$, at which the lattice reaches a soft mode instability in the absence of deviatoric deformation. For each sampled value of $\bar{\mathcal E}$, and along each sampled deviatoric stretch angle $\theta$,  the superimposed deviatoric strain is varied such that its magnitude $\gamma$ varies over the range  $0 \le \gamma \le \gamma^c$, the upper limit $ \gamma^c$ being the critical value of the stretch at the given value of $\bar{\mathcal E}$ and $\theta$. Owing to the $\mathcal C_{6v}$ symmetry of graphene, the deviatoric stretch angle $\theta$ needs to be sampled only over the range $0 \le \theta \le \pi/6$. Further, we consider only $N =2$ deformation paths in the BZ, corresponding to the zigzag direction $\theta =0$,  and the armchair direction $\theta=\pi/6$. Later, we will show that only two sampling directions suffice for an accurate angular interpolation of the failure function.   
 \subsection{Representation and interpretation of the data} 
From the lattice-dynamical stability analysis, we obtain the limiting values, each in the form a triplet $\{\bar{\mathcal E},\, \gamma^c,\, \theta\}$,  representing the critical value of the deviatoric strain magnitude --- $\gamma^c$ --- as a function of mean hydrostatic strain $
%
\bar{\mathcal E}$ for both the sampled deformation paths: $\theta=0$ (shown in Fig.~(\ref{sampling}-b)) and $\theta=\pi/6$ (shown in Fig.~(\ref{sampling}-c)). \\ \\
\begin{minipage}{\linewidth}
\centering
\includegraphics[scale=0.55]{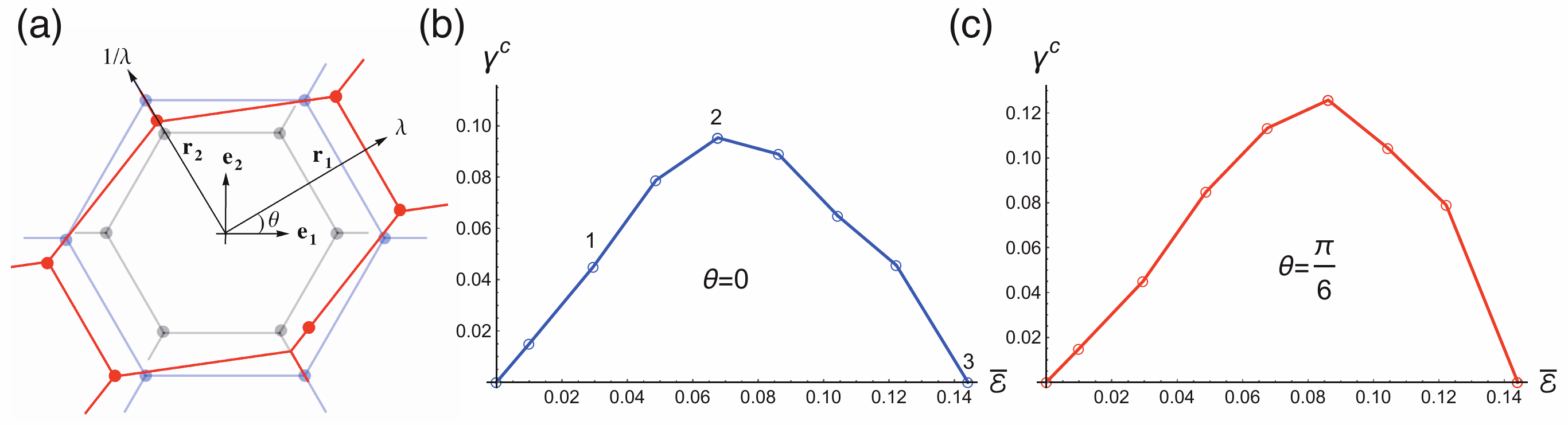}
\captionof{figure}{\small (a) Notation scheme illustrating  deformation of the graphene lattice: the undeformed lattice (faint black) is  first subjected to a uniform dilatation $\mathbf U_1 = J^{1/2} \mathbf I$ (resulting intermediate configuration shown in faint blue), and then to an isochoric shape-changing deformation $\mathbf U_2 = \lambda \mathbf r_1 \otimes \mathbf r_1 + \lambda^{-1} \mathbf r_2 \otimes \mathbf r_2$ (resulting final configuration shown in dark red). (b) 
Cartesian plot of $\gamma^c$ as a function of $\bar{\mathcal E}$ for the sampled orientation $ \theta =0$, which corresponds to the zigzag direction. (c) Cartesian plot of $\gamma^c$ as a function of $\bar{\mathcal E}$ for the sampled orientation $\theta=\pi/6$, which corresponds to the armchair direction. }
\label{sampling}
\end{minipage}
\\ \\
Shown in Fig.~(\ref{phonons}) are the phonon dispersion relations of graphene calculated with DFPT at certain critical deformed states, indicated as `1', `2' and `3', along the sampled deformation path $\theta=0$; these particular critical states of deformation are those labeled on Fig.~(\ref{sampling}-b). Depending upon the level of mean hydrostatic strain, a critical phonon state might be associated with buckling, as in `1', for which the long-wave ZA phonon slope at $\Gamma$ vanishes; with an elastic (material) instability, as in `2', for which the long-wave LA slope at $\Gamma$ vanishes; or with a short-wave material instability as in `3', where the frequency of the TA phonon vanishes at $K$. Thus, at different states of deformation, different kinds of mechanisms govern lattice instability in a graphene sheet. The proposed limit surface is a smoothed envelope of all the possible lattice-instabilities, as we will elaborate further in Sec.(\ref{envelop}). \\ \\

\begin{minipage}{\linewidth}
\centering
\includegraphics[scale=0.165]{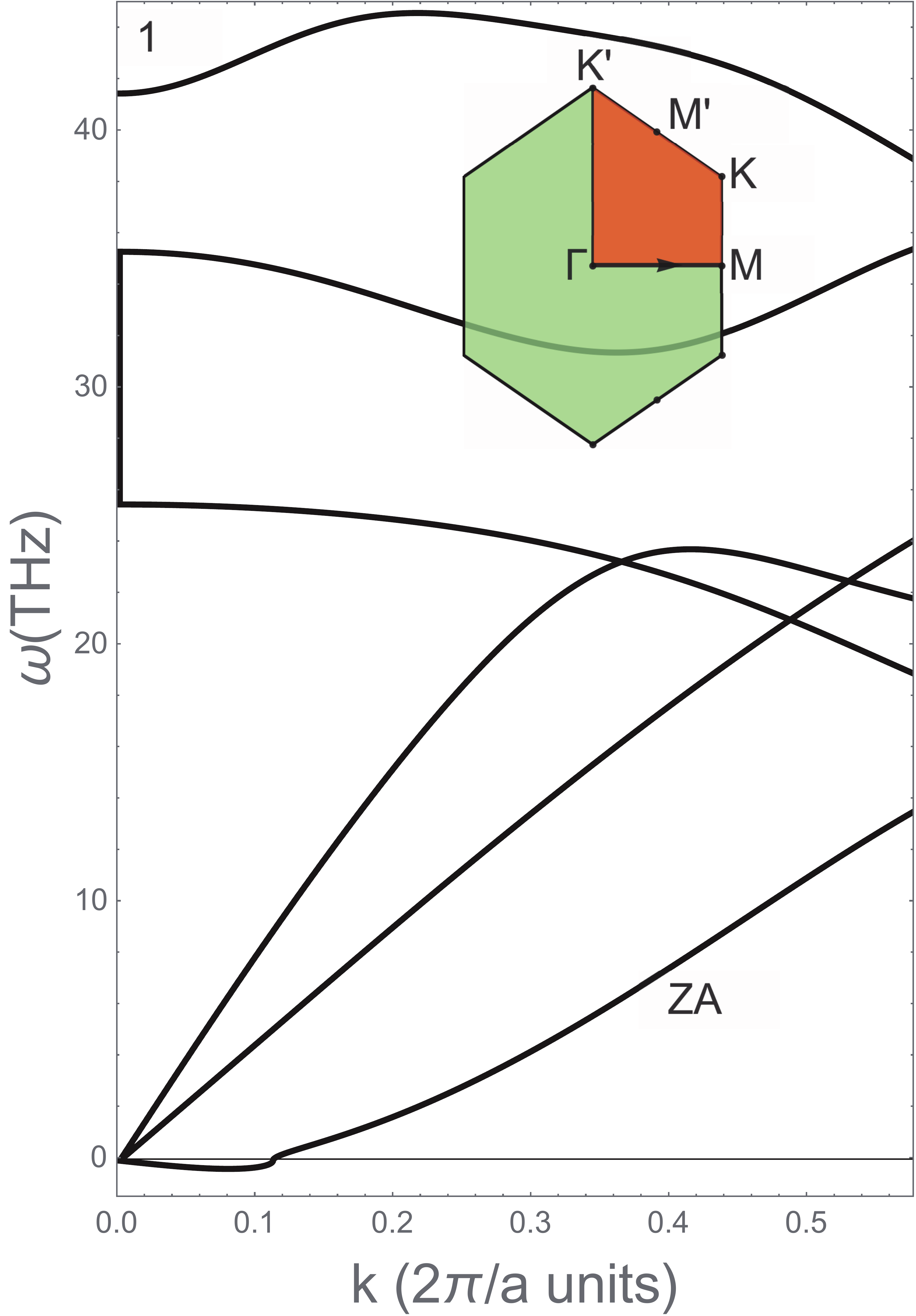} \includegraphics[scale=0.165]{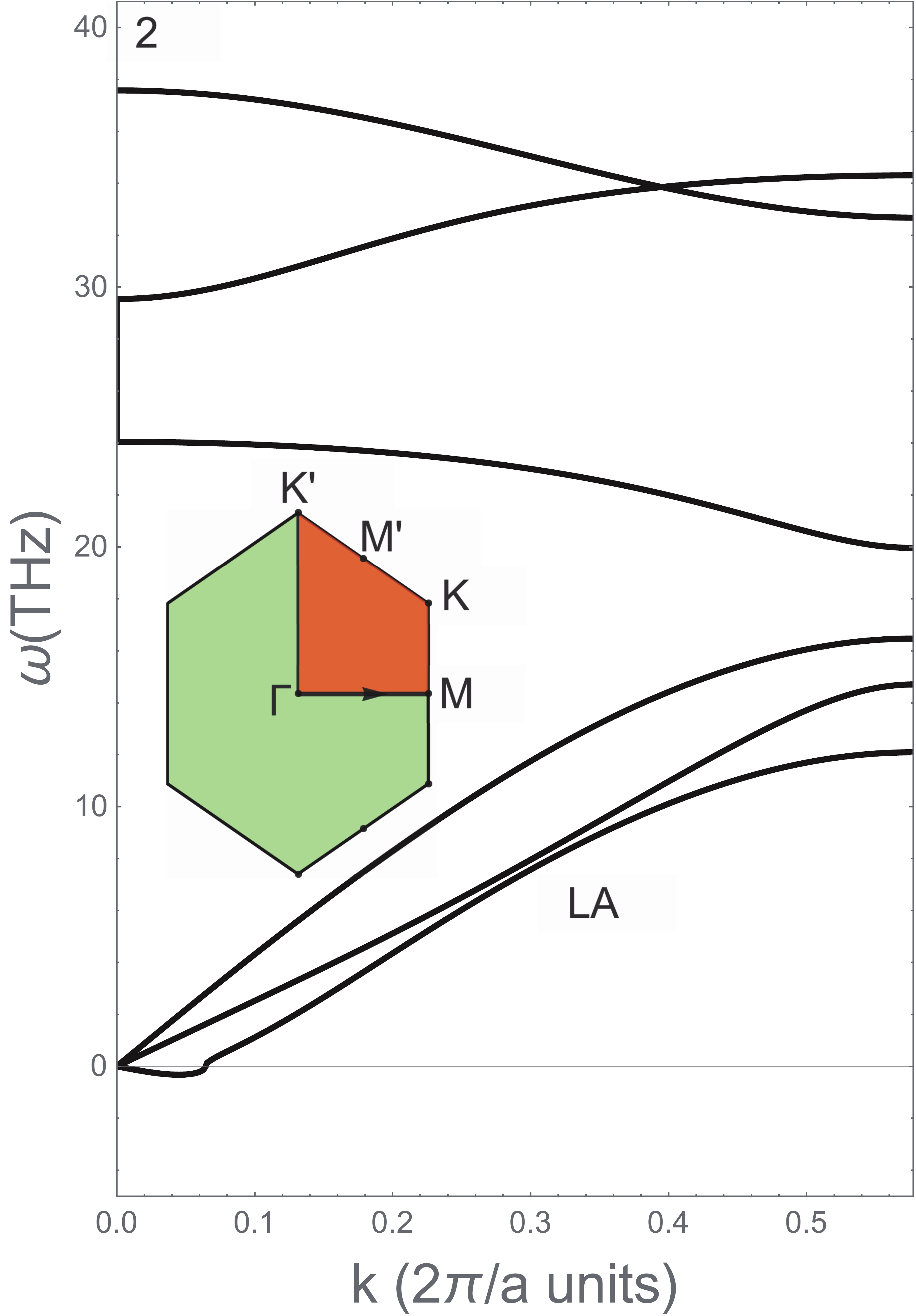} \includegraphics[scale=0.165]{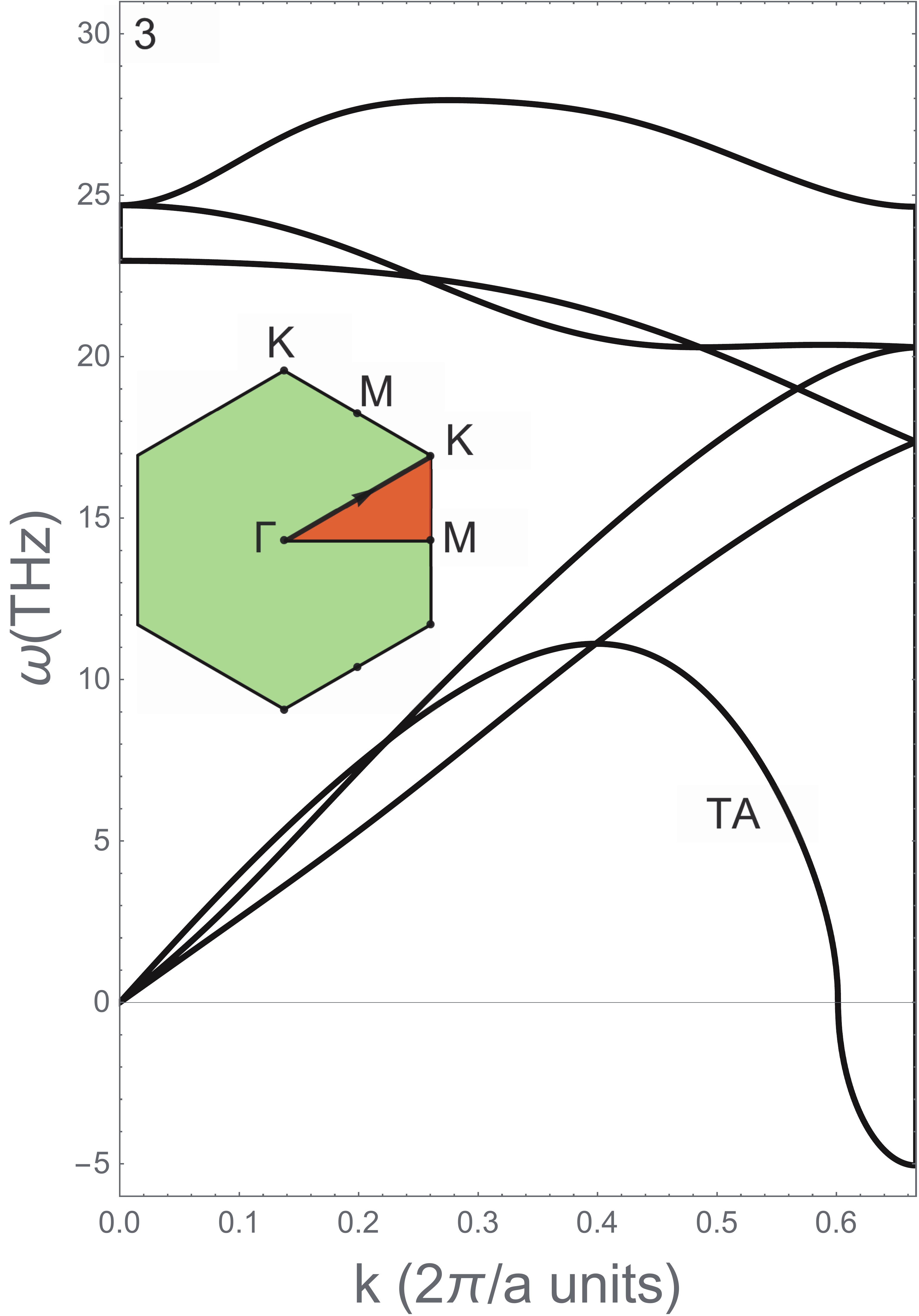}
\captionof{figure}{ Phonon dispersions of graphene at certain critical states of deformation, marked as `1', `2' and `3' in Fig.~(\ref{sampling}b), along the armchair deformation path ($\theta =0$). Also shown are the deformed Brillouin zone (green polygon), and the irreducible wedge (red polygon). The sampled direction in each case is shown by an arrow. The labels `ZA', `LA' and `TA' show the branch that goes unstable in the three cases. 
}
\label{phonons}
\end{minipage}
\subsection{From discrete dataset to continuum lattice-stability limit function via interpolation}
Construction of a continuum-level limit function involves interpolating the discrete dataset of the preceding section in a manner that is consistent with the material symmetry ($\mathcal C_{6v}$) of graphene. This task is accomplished by interpolating the critical magnitude of deviatoric strain $\gamma^c$ in terms of two of the symmetry-invariants of the strain tensor: $\mathcal E$ and $\Theta(\theta)$. The interpolation is performed in two steps: first, interpolation in terms of $\bar{\mathcal E}$, called radial interpolation, and second, interpolation in terms of $\Theta(\theta)$, called angular interpolation.
 \subsubsection{Radial interpolation (interpolation in terms of $\bar{\mathcal E}$)}
Along each sampled deformation path $\Theta =\Theta_n$, we express $\gamma^c$ by a continuous function $\mathcal R_n (\bar{\mathcal E} ) $ --- called a radial interpolation function. Each such function is obtained by fitting the sampled datapoints (plotted in Fig.~(\ref{radyif}-a)), which are in the form of a doublet $\{\bar{\mathcal E},\ \gamma^c \}$ and denote the limiting strains along a sampled deformation path, to a polynomial function in $\bar{\mathcal E}$ of the form:
\begin{equation}
 \begin{aligned}[b]
\gamma^c (\bar{\mathcal E}, \Theta = \Theta_n) & \equiv  \mathcal  R_n(\bar{\mathcal E}/\bar{\mathcal E}_{\mathrm{c}})     \\
 & = \beta_n^1 (\bar{\mathcal E}/\bar{\mathcal E}_{\mathrm{c}} ) + \beta_n^2 (\bar{\mathcal E}/\bar{\mathcal E}_{\mathrm{c}} )^2 + \beta_n^3 (\bar{\mathcal E}/\bar{\mathcal E}_{\mathrm{c}} )^3 + ... +\beta_n^{M}(\bar{\mathcal E}/\bar{\mathcal E}_{\mathrm{c}} )^{M},
 \label{majorE}
 \end{aligned}
 \end{equation}
 where $ \bar{\mathcal E}_{\mathrm{c}} = \ln J_{\text{max}} \doteq 0.144$ is the mean hydrostatic strain at which an incipient lattice instability emerges without any deviatoric strain.
The radial interpolation functions thus obtained, corresponding to the two sampled deformation paths, are shown in Fig.~(\ref{radyif}-b).
%
\\
\begin{minipage}{\linewidth}
\centering
\includegraphics[scale=0.6]{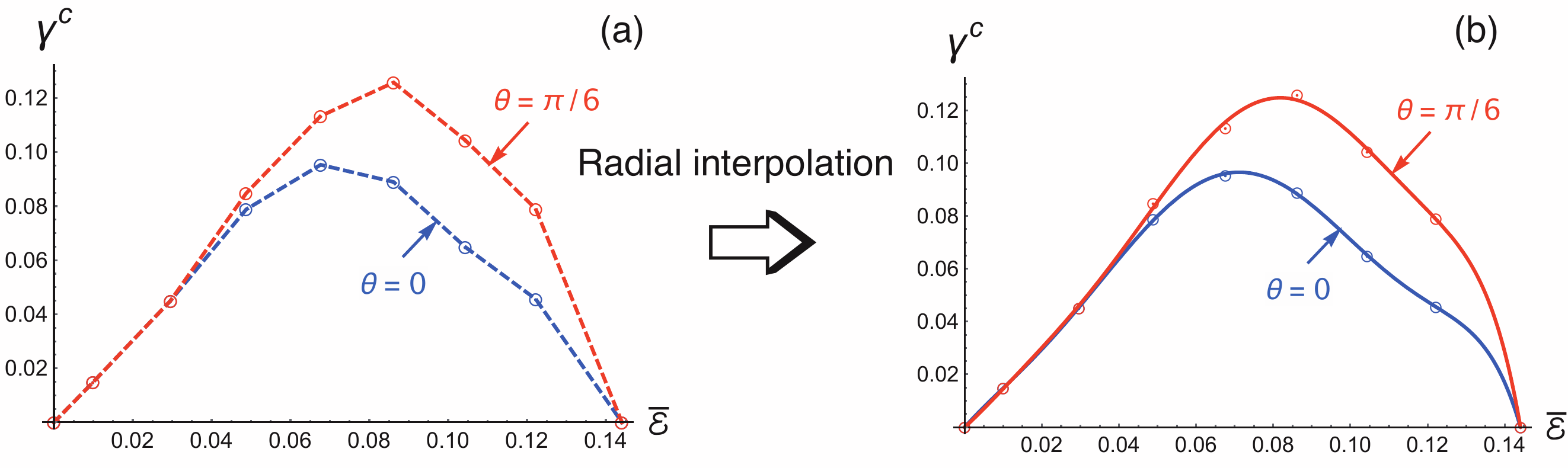} 
\captionof{figure}{\small (a) Sampled valued of $\gamma^{c}$ along two deformation paths: $\theta =0$ and $\theta=\pi/6$ (straight-line segments connect data  points).  
(b) $7^{th}$-order polynomial interpolation functions for $\gamma^c$ along the armchair and zigzag directions ---  obtained by fitting functions described by equation~\ref{majorE}.  }
%
\label{radyif}
\end{minipage}
\subsubsection{Angular interpolation}
The angular interpolation scheme involves approximating the failure surface $\gamma^c (\bar{\mathcal E}, \Theta)$ by a  weighted sum of the radial interpolation functions --- with the weight functions being the angular shape functions along the sampled deformation paths. The following representation for the angular shape function $\mathcal K_n (\Theta)$ conveniently satisfies the requirements that failure function is smooth, satisfies all the point-group symmetries of the lattice, and possesses $2\pi$ periodicity with respect to $\Theta$:
 \begin{equation}
 \mathcal K_{n}(\Theta)= \alpha_n^0 + \alpha_n^1 \cos(\Theta) +\alpha_n^2 \cos(2\Theta) +...;
 \label{a} \end{equation}
 where the number of terms in the expansion is given by the number of deformation paths ($N$) being sampled, and the corresponding coefficients $\alpha_n$ are determined using the following condition:
 \begin{equation}
\mathcal K_m(\Theta_n) = \begin{cases} 0 ; & \Theta_n \ne \Theta_m  \\
                    1; & \Theta_n = \Theta_m.  \end{cases}
                    \label{shapefunctions}
\end{equation}
Since the angular variation of the limit function is quite smooth, sampling along only two deformation paths, i.e., $N=2$, suffices for accurate interpolation of the limit surface (see Fig.~(\ref{Validate}) in Sec.(\ref{validation})). The corresponding two shape-functions are given as:
\begin{eqnarray}
\mathcal K_1(\Theta) = \frac{1}{2}(1+  \cos ( \Theta )),\\
   \mathcal K_2(\Theta) =  \frac{1}{2}(1- \cos ( \Theta )).
  \end{eqnarray}
Following the above interpolation scheme, $\gamma^c$ is generically expressed as:
\begin{equation}
\gamma^c(\bar{\mathcal E}, \Theta(\theta))=\sum_{n=1}^{n= N} \mathcal R_n(\bar{\mathcal E}/\bar{\mathcal E}_{c}) \mathcal K_n(\Theta(\theta)),
\label{sch}
\end{equation}
where $N=2$ is the total number of deformation paths. 
Substituting from equation~(\ref{majorE}) and equation~(\ref{a}) into  equation~(\ref{sch}), we obtain the  generic form of the lattice stability function as:
 \begin{equation}
 \gamma^c(\bar{\mathcal E}, \Theta(\theta))=  \sum_{n=1}^{n=N} \sum_{{m=1}}^{M} \gamma_{mn}  (\bar{\mathcal E}/\bar{\mathcal E}_{c})^m \cos(n-1)\Theta(\theta);
 \label{yseqn1}
 \end{equation}
 The various coefficients appearing in the expression of the failure function have been tabulated in the Tab.~(\ref{Tabcoeffs}). \\
\begin{table}[htbp!]
\hspace*{-1cm}
\begin{tabular}{c|ccccccc}
\toprule[1.5pt] 
$\gamma_{mn}$  $m \rightarrow$ \\ $ \ \ \ \ \  n \downarrow$& $\bar{\mathcal E}/\bar{\mathcal E}_{c}$ &  $(\bar{\mathcal E}/\bar{\mathcal E}_{c})^2$ &$(\bar{\mathcal E}/\bar{\mathcal E}_{c})^3$  & $(\bar{\mathcal E}/\bar{\mathcal E}_{c})^4$ &$(\bar{\mathcal E}/\bar{\mathcal E}_{c})^5$ &$(\bar{\mathcal E}/\bar{\mathcal E}_{c})^6$ & $(\bar{\mathcal E}/\bar{\mathcal E}_{c})^7$ \\
\hline 
\hline
\\
1&   \hspace*{0.2cm}  0.21561 \hspace*{0.2cm} & \hspace*{0.2cm} - 0.0403958 \hspace*{0.2cm}  &  \hspace*{0.2cm}  - 0.327806 \hspace*{0.2cm}  &  \hspace*{0.2cm}  7.51334   \hspace*{0.2cm}  & \hspace*{0.2cm}  - 24.8089 \hspace*{0.2cm}  & \hspace*{0.2cm} $ 28.3255 $  \hspace*{0.2cm} & \hspace*{0.2cm}$ - 10.8837$\hspace*{0.2cm} \\ \\
\hline
\\
$\cos \Theta$ & 0.0251542 & - 0.583471    &4.55324 &  -16.0798 & 26.6759& - 20.8193 & 6.22982 \\ \\
\hline
\\
\bottomrule[1.5pt]
\end{tabular}
\caption{ Coefficients $\gamma_{mn}$ in the expression of the failure surface given by equation~(\ref{yseqn1}); the basis function for each coefficient $\gamma_{mn}$ is the product of the $m^{th}$ column of the top-most row with the $n^{th}$ row of the left-most column.}
\label{Tabcoeffs}
\end{table}
%
\\
The $0 \le \theta \le \pi/2$ quadrant of the instability surface resulting from the interpolation scheme is shown in Fig.~(\ref{radyifII}-b). \\
 \begin{minipage}{\linewidth}
\centering
\includegraphics[scale=0.55]{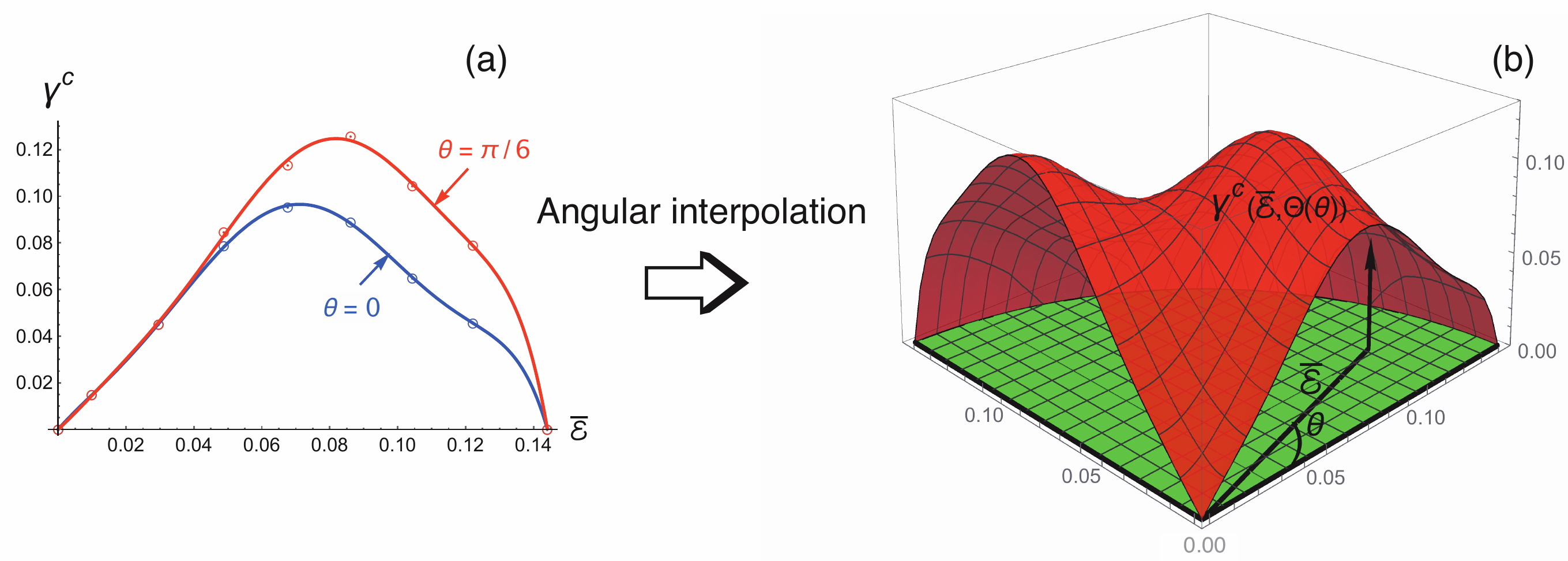}
%
\captionof{figure}{\small (a) The radial interpolation of the lattice-stability limits along the two sampled directions, $\theta=0$ and $\theta=\pi/2$.  (b) A $90^o$ quadrant cut out from the limit surface.  From below, the limit surface is bounded by the zero surface shown in green; while from above the failure surface is bounded by the critical surface $\gamma = \gamma^c(\bar{\mathcal E}, \Theta(\theta))$, shown in red. The directions $\theta=0$ and $\theta=\pi/2$ correspond to the zigzag and armchair directions, respectively.}
\label{radyifII}
\end{minipage}
\\ \\
\subsection{Limit surface as a smoothed envelope of various possible instabilities}
\label{envelop}
Different kinds of strength-limiting mechanisms lead to instability of the graphene lattice under different modes of deformation. The limit surface is a smoothed representation of the envelope of all possible lattice instabilities: long wavelength as well as short wavelength, and structural as well as material failures. 

Graphene, being a strictly two-D structure, has an exceedingly small bending rigidity.  Therefore, 
when subjected to an in-plane deformation, a graphene sheet can be brought to a state of incipient instability via one of  two mechanisms: 
\begin{itemize}
\item a material instability in tension, which arises when the major principal strain $\mathcal E_{\text{max}}$ (maximum strain over all material directions) reaches a limiting value, while the minor principal strain $\mathcal E_{\text{min}}$ (minimum strain over all material directions) has not reached the critical value in compression so that both the Cauchy principal stress components are non-negative, i.e., $\mathcal S_{\text{max}}\ge \mathcal S_{\text{min}} \ge 0$. This material instability can be either of the two types: long-wave (also called elastic) or a short-wave (also called soft mode)
or
\item  a structural instability via out-of-plane buckling, which arises when the minor principal strain $\mathcal E_{\text{min}}$ (minimum strain over all material directions) in the material reaches the critical value in compression such that  one of the principal stress components $\mathcal S_{\text{max}}$ or $\mathcal S_{\text{min}}$ becomes incipiently compressive ($<0$), while the major principal strain has still not reached the limiting tensile strain. 
\end{itemize}

The two principal strains, in terms of $\bar{\mathcal E}$ and $\gamma$, are given by $\mathcal E_{\text{max}} = \bar{\mathcal E} + \gamma$ and $\mathcal E_{\text{min}} = \bar{\mathcal E} - \gamma$. Thus, at a fixed equi-biaxial strain $\bar{\mathcal E}$, with a superposed, increasing deviatoric strain $\gamma$, the major principal strain monotonically increases while the minor principal strain monotonically decreases. Now, whether a deviatoric strain $\gamma$ superposed on an equi-biaxial strain $\bar{\mathcal E}$ results in a material instability or a structural instability depends on the magnitude of $\bar{\mathcal E}$, as schematically illustrated in Fig.~(\ref{mechanisms}). For example, when $\bar{\mathcal E}$ is small, the minor value $\mathcal E_{\text{min}}$ reaches the critical value in compression, which is $ \sim -\nu_{2D} \mathcal E_{\text{max}}$,  before the major principal value $\mathcal E_{\text{max}}$ reaches the limiting tensile strain. Thus, at small equi-biaxial strain, the magnitude of the deviatoric strain $\gamma$ is essentially limited by a buckling instability. This is also shown in the phonon dispersion relation of Fig.~(\ref{phonons}-1). Interestingly, when $\bar{\mathcal E} = 0$, the membrane is unstable in buckling as soon as any deviatoric strain is applied; thus the limiting value of $\gamma^c |_{\bar{\mathcal E} \rightarrow 0} =0$ $\forall \ \theta$.\\
\hspace*{0.5cm} On the other hand, when $\bar{\mathcal E}$ is large, $\mathcal E_{\text{max}}$ reaches the limiting tensile strain before $\mathcal E_{\text{min}}$ reaches the critical value in compression, and the failure is invariably a material failure (fracture). Further, as $\bar{\mathcal E}$ approaches closer to the critical value $\bar{\mathcal E}_{c}$, the material starts to fail essentially in dilatation via a short-wave instability, as shown in Fig.~(\ref{phonons}-3). Thus, $\gamma^c |_{\bar{\mathcal E} \rightarrow \bar{\mathcal E}_{c}} \rightarrow 0\ \forall\  \theta$.\\
\begin{minipage}{\linewidth}
\centering
\includegraphics[scale=0.41]{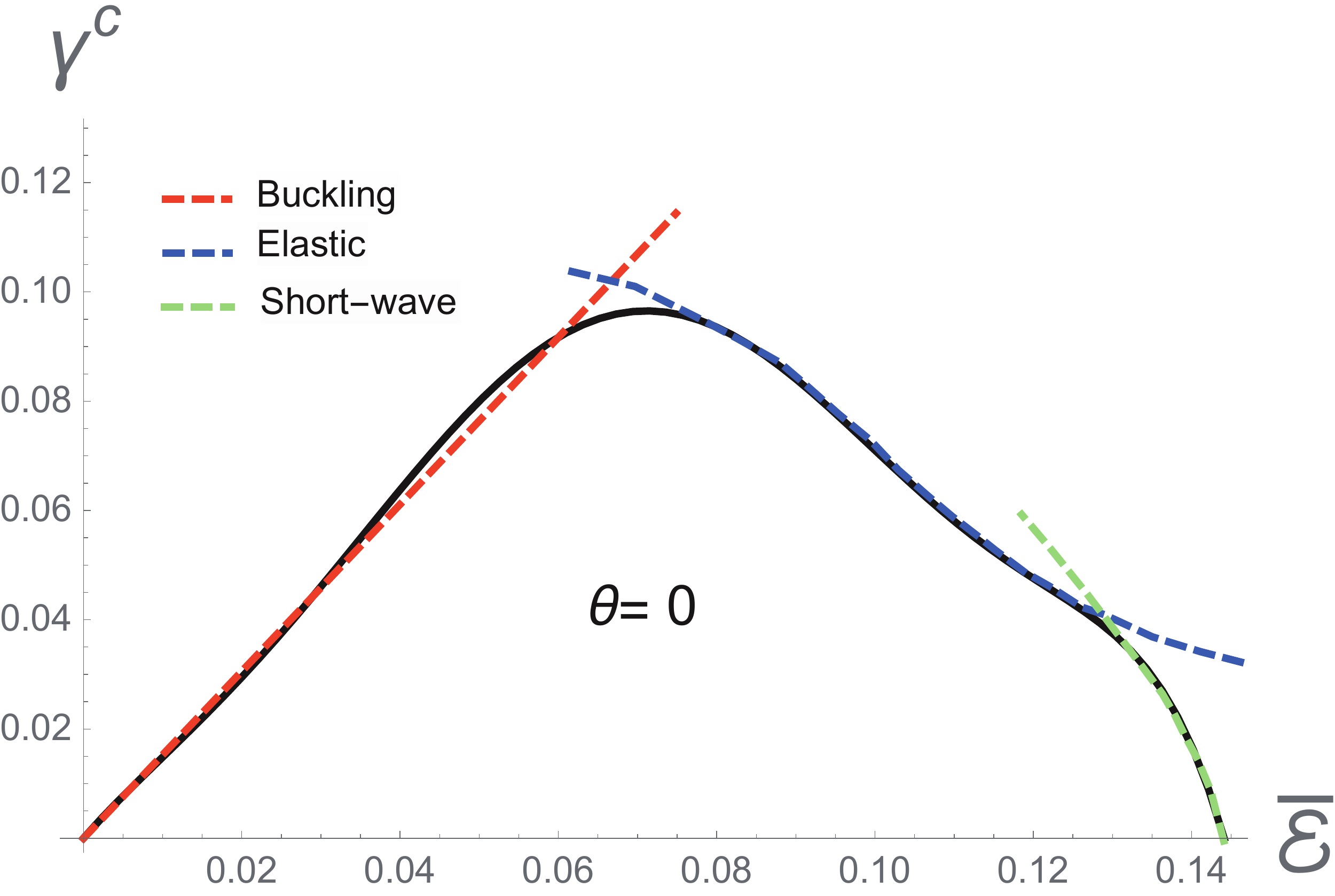}
\captionof{figure}{\small Schematic illustration showing the lattice-stability plot as a smoothed representation of the envelope of the various possible lattice instabilities.}
\label{mechanisms}
\end{minipage} \\ \\
%
%

\section{Lattice-stability limit surface in stress space}
\label{straintostressspace}
Often, it is useful to describe the lattice-stability limits in terms of stress. In the present case, although the limiting conditions are explicitly obtained in terms of the invariants of strain, the relationship can be numerically mapped to its counterpart in the stress space via a constitutive function appropriately describing the stress-strain response in graphene over the entire range of deformation up to the elastic stability limit. In a previous work \cite{kumar2014on}, we derived a large-deformation hyperelastic constitutive response function for graphene based on \textit{ab initio} calculations --- which we briefly describe in the following.\\
The constitutive response function is derived within the framework of nonlinear hyperelasticity and invariant-based representation theory. Our formulation employs the symmetry-invariants of the log strain measure $\mathbf E^{(0)}$ --- $\epsilon_a, \gamma_i^2, \text{and } \gamma_{\theta}^3$ --- to express the dependence of  the free energy per unit reference area  $\psi$ on the state of strain $\mathbf E^{(0)}$, i.e., $\psi = \hat \psi(\epsilon_a, \gamma_i^2,  \gamma_{\theta}^3) $ (see Kumar \& Parks \cite{kumar2014on} for details). The strain energy density function 
$\hat \psi(\epsilon_a, \gamma_i^2,  \gamma_{\theta}^3)$ accepts the additive decomposition: $
\psi=  \hat \psi^{\mathrm{Dil}}(\epsilon_a) +\hat \psi^{\mathrm{Dev}} (\epsilon_a, \gamma_i^2 , \gamma_{\theta}^3)$.
The term $\hat \psi^{\mathrm{Dil}}$ --- the energetic response under pure dilation --- is described by a function based on the universal binding energy relation \cite{rose1983universal, rose1981universal}: 
\begin{equation}
 \hat \psi^{\text{Dil}}(\epsilon_a) = \psi_o\left[1 - (1+ \alpha \epsilon_a) \exp (-\alpha \epsilon_a)\right].
 \end{equation}
The term $\hat \psi^{\mathrm{Dev}}$ --- the energetic response under an isochoric deformation --- is given by a simple additive form:
\begin{equation}
\hat \psi^{\mathrm{Dev}}(\epsilon_a, \gamma_i^2 , \gamma_{\theta}^3) =  \frac{1}{2} \mu (\epsilon_a) \gamma_i^2 +  \frac{1}{8}\eta(\epsilon_a) \gamma_{\theta}^3 ,
\end{equation}
where $\mu (\epsilon_a) = \mu_0 - \mu_1 e^{\beta \epsilon_a}$ is a second order elastic constant, and $ \eta (\epsilon_a)  = \eta_0  - \eta_1 \epsilon_a^2$ is a third-order elastic constant. The coefficients are determined from fits to \textit{ab initio} energies for a set of deformed configurations (see Tab.[\ref{coeffs}]). 
\begin{table}[htbp!]
\begin{center}
\begin{tabular}{c c c c c c c c c c }
\toprule[1.5pt]
& $\psi_o$(J/m$^2$) & $\alpha$ &$\mu_0 $(N/m) & $\mu_1$(N/m) & $\beta$& $\eta_0$(N/m) & $\eta_1$(N/m) \\
\hline 
\hline
 & & & & & & & \\
LDA & 116.43& 1.38 & 172.18 & 27.03 & 5.16  &93.17 &4408.76 \\
\bottomrule[1.65pt]
\end{tabular}
\end{center}
\caption{\small Coefficients in the expression of strain energy function $\hat \psi (\epsilon_a, \gamma_i^2 , \gamma_{\theta}^3)$.}
\label{coeffs}
\end{table}
Differentiation of $\psi$ with respect to $\mathbf E^{(0)}$ gives the work-conjugate stress tensor, i.e., $\mathbf T^{(0)} = \partial \psi / \partial \mathbf E^{(0)}$. 
\begin{multline}
\mathbf T^{(0)}(\epsilon_a, \gamma_i^2, \gamma_{\theta}^3)= \frac{\partial \psi }{\partial \mathbf E^{(0)}} \\
= \bigg \lgroup \frac{\partial \hat \psi^{\mathrm{Dil}}(\epsilon_a)}{ \partial \epsilon_a} +\frac{\partial \hat \psi^{\mathrm{Dev}}(\epsilon_a, \gamma_i^2, \gamma_{\theta}^3)}{ \partial \epsilon_a}\bigg \rgroup \mathbf I   + 4 \frac{\partial \hat \psi^{\mathrm{Dev}}(\epsilon_a, \gamma_i^2, \gamma_{\theta}^3)}{ \partial (\gamma_i^2)}  \mathbf E^{(0)}_0 + \frac{\partial \hat \psi^{\mathrm{Dev}}(\epsilon_a, \gamma_i^2, \gamma_{\theta}^3)}{ \partial (\gamma_{\theta}^3)}\mathbf S_{\mathbf E^{(0)}_0} \\
=\bigg \lgroup \psi_o \alpha^2 \epsilon_a \exp(-\alpha\epsilon_a)+ \frac{1}{2}\mu^{\prime}(\epsilon_a)\gamma_i^2+ \frac{1}{8} \eta^{\prime}(\epsilon_a) \gamma_{\theta}^3\bigg \rgroup \mathbf I + 2 \mu(\epsilon_a) \mathbf E^{(0)}_0 + 
\frac{1}{8}\eta(\epsilon_a)\mathbf S_{\mathbf E^{(0)}_0},
\label{wcstress}
\end{multline}
and the Cauchy stress $\pmb \sigma$ is given as (see Ogden \cite{ogden1997non})
\begin{equation}
\pmb \sigma = \frac{1}{J} \mathbf F \mathbf T^{(2)}\mathbf F^T = 2 \frac{1}{J} \mathbf F\left[\frac{\partial \mathbf E^{(0)} }{\partial \mathbf C \ \ \ } \right] \mathbf T^{(0)}\mathbf F^T ,
\label{cauchy}
\end{equation}
where $\mathbf C=\mathbf U^2$ denotes the right Cauchy-Green stretch tensor.
Employing equation~(\ref{wcstress}) and equation~(\ref{cauchy}), we numerically map the contours in the strain plane to stress-plane. Once the Cauchy stress has been determined, we can obtain the principal stresses from the spectral representation of $\pmb \sigma $: 
\begin{equation}
\pmb \sigma = \mathcal S_{\text{max}} \mathbf s_1 \otimes \mathbf s_1 + \mathcal S_{\text{min}} \mathbf s_2 \otimes \mathbf s_2,
\end{equation}
where $\mathcal S_{\text{max}}$, and $\mathcal S_{\text{min}}$ are the principal Cauchy stress components; and $\mathbf s_1 = \cos \phi\, \mathbf e_1 + \sin \phi\, \mathbf e_2 $ and  $\mathbf s_2 = -\sin \phi\, \mathbf e_1 + \cos \phi\, \mathbf e_2 $ are the corresponding principal directions. The orientation $\phi$ is measured from a fixed material axis aligned with the zigzag direction of the graphene lattice.\\
Employing the above transformation, we obtain the lattice-stability plot in stress space as a  functional relation between the components of the functional bases comprising the maximum shear stress $\tau_{\text{max}} =  (\mathcal S_{\text{max}}-\mathcal S_{\text{min}})/2 $, the mean hydrostatic Cauchy stress $\bar {\sigma} = (\mathcal S_{\text{max}}+\mathcal S_{\text{min}})/2$, and the symmetry-reduced deviatoric stress direction $\Phi=\arccos(\cos(6\phi))$. The limit function--- in terms of $\tau_{\text{max}}$, $\bar {\sigma}$, and $\Phi(\phi)$ --- can be expressed in a form similar to the limit function in strain space, i.e.,
\begin{equation}
\mathcal T(\pmb \sigma) \equiv \tau_{\text{max}}^c (\bar{\sigma}, \Phi(\phi) ) - \tau_{\text{max}} =0,\ \tau_{\text{max}} \ge 0;
\label{tmax}
\end{equation}
such that if, at a given state of deformation, $\mathcal T (\pmb \sigma) >0 $, then the material is stable with respect to  incremental perturbations of all wavelengths; otherwise the instantaneous configuration of the lattice is unstable under some incremental perturbation. \\ \\
The surfaces described by equation~(\ref{tmax}) constitute the lattice-stability limit surface in the space of symmetry-invariants of the Cauchy stress tensor, as shown in Fig.~(\ref{straintostress}-b).
\\
\begin{minipage}{\linewidth}
\centering
\includegraphics[scale=0.62]{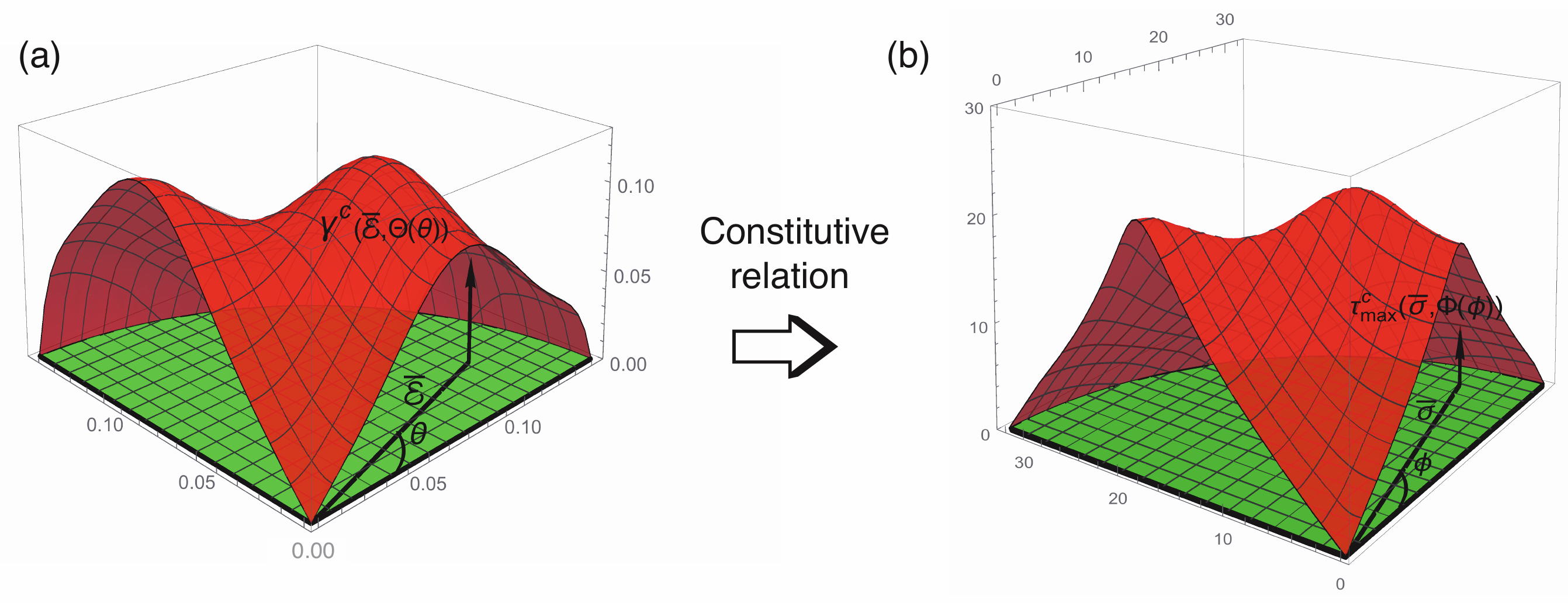}
\captionof{figure}{\small Numerical mapping of the instability surface from strain space (shown in (a)) to stress space (shown in (b)) via the hyperelastic constitutive relation of equation(\ref{wcstress}). The axes in stress space are in units of N/m. The directions $\phi=0$ and $\phi=\pi/2$ correspond to the zigzag and armchair directions, respectively.}
\label{straintostress}
\end{minipage} 

\section{Validation}
\label{validation}
The limit surface  --- given by equation~(\ref{yseqn1}) 
 --- contains information of all possible lattice-instabilities that can be triggered by a homogeneous  deformation. 
 Specifically, once this failure surface is determined,  we can assess the stability of graphene subject to a biaxial state of strain --- $ \mathcal E_{11}^{\theta}\, \mathbf r_1 \otimes \mathbf r_1 +\mathcal E_{22}^{\theta}\, \mathbf r_2 \otimes \mathbf r_2 $ --- for any biaxiality (the $\mathcal E_{11}^{\theta}$ to $\mathcal E_{22}^{\theta}$ ratio) and any deviatoric stretch angle $\theta$. For such deformations, the principal strains are directly given by the strain components $\mathcal E_{11}^{\theta}$ and $\mathcal E_{22}^{\theta}$ as $\mathcal E_{\text{max}} = \text{max}[\mathcal E_{11}^{\theta}, \mathcal E_{22}^{\theta}]$ and $\mathcal E_{\text{min}} = \text{min} [\mathcal E_{11}^{\theta}, \mathcal E_{22}^{\theta}]$. Substituting a particular value of $\theta$ in the limit surface, we can obtain the 2D radial section of the limit surface corresponding to that particular value of $\theta$. For example, Fig.~(\ref{cutplane}) shows the 2D radial sections of the limit surface corresponding to $\theta =0$ and $\theta=\pi/6$, denoted by $\gamma^c\rvert_{\theta=0} -\gamma=0$ and $\gamma^c\rvert_{\theta=\pi/6} -\gamma=0$, respectively. These contours characterize the stability of a graphene sheet subjected to a homogeneous biaxial state of strain referred to the set of axes defined by $\theta=0$ and $\theta=\pi/6$, i.e., $\mathbf r_1-\mathbf r_2$ are aligned with $\mathbf e_1-\mathbf e_2$ (see Fig.~(\ref{sampling}-a)). 
 
On this contour, we have indicated points corresponding to certain special homogenous deformation cases such as uniaxial strain and uniaxial stress along the armchair and zigzag directions : 
\begin{enumerate}
\item The point indicated as $\blacktriangledown_z$, given by the intersection of the line $\gamma=\bar{\mathcal E}$ with the contour $\gamma^c \rvert_{\theta =0} - \gamma =0$, denotes the uniaxial strain along the zigzag direction.
\item The point indicated as $\blacktriangledown_a$, given by the intersection of the line $\gamma=\bar{\mathcal E}$ with the contour $\gamma^c \rvert_{\theta =\pi/6} - \gamma =0$, denotes the uniaxial strain along the armchair direction.
\item The point indicated as $\blacktriangle_a$, given by the intersection of the line $\gamma=\left(\frac{1+\nu_{2D}}{1-\nu_{2D}} \right)\bar{\mathcal E}$ with the contour $\gamma^c \rvert_{\theta =0} - \gamma =0$, denotes elastic failure under uniaxial tension in the armchair direction.
\item Similarly, the point marked $\blacktriangle_z$, given by the intersection of the line $\gamma=\left(\frac{1+\nu_{2D}}{1-\nu_{2D}} \right)\bar{\mathcal E}$ with the contour $\gamma^c \rvert_{\theta =\pi/6} - \gamma =0$, denotes the uniaxial tension in the zigzag direction.
\item The point indicated by $\blacksquare$, given by the intersection of the line $\gamma=0$ with a contour $\gamma^c \rvert_{\theta} - \gamma =0$ ($\forall \ \theta$), denotes the limiting strain under equi-biaxial strain.
\end{enumerate}
 Following the above-mentioned procedure, it is possible to obtain the limiting contour corresponding to any directionality $\theta$, and deduce the corresponding stability-limiting values of stress/strain in uniaxial strain/stress.\\\\
\begin{minipage}{\linewidth}
\centering
\includegraphics[scale=0.6]{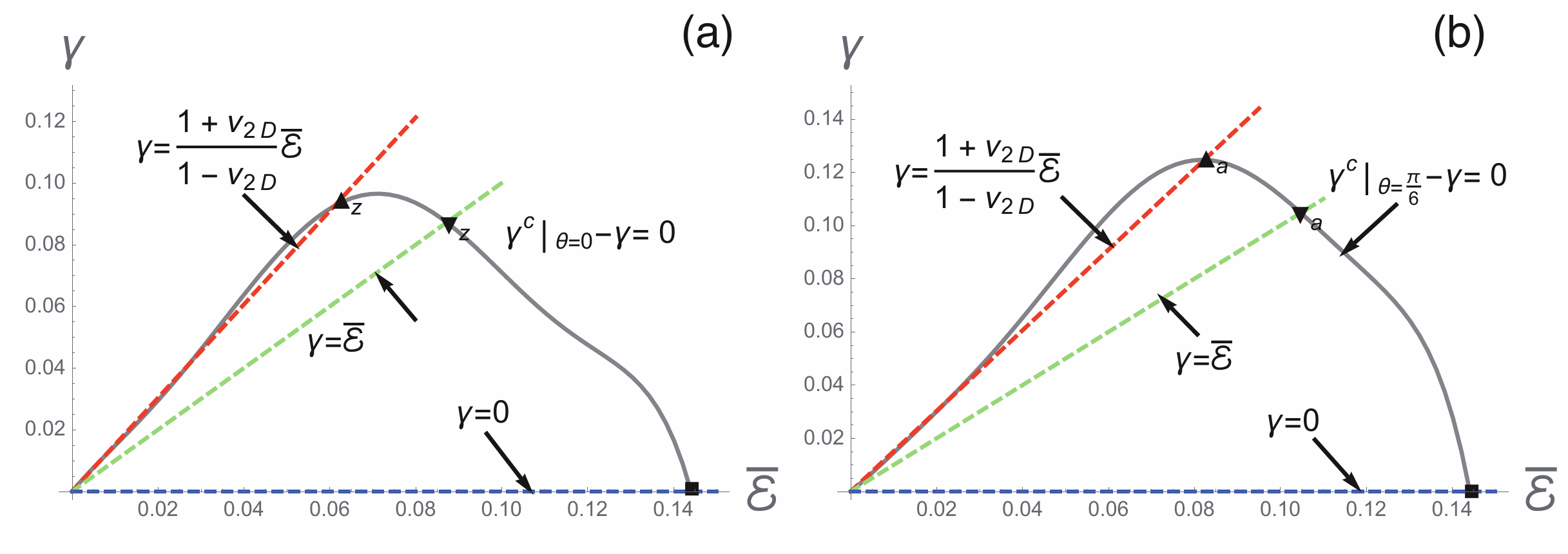}
\captionof{figure}{\small (a) Cross-section of the lattice-stability limit surface corresponding to $\theta=0$. Critical points correspond to  special homogeneous deformed state such as uniaxial stress and uniaxial strain in the zigzag direction and equi-biaxial strain are indicated. (b) Cross-section of the lattice-stability limit surface corresponding to $\theta=\pi/6$. Critical points corresponding to special homogeneously deformed states such as uniaxial stress and uniaxial strain in the armchair direction and equi-biaxial strain are indicated. }
\label{cutplane}
\end{minipage} \\ \\
\subsection{Comparison between predicted response and values calculated from DFPT}
\label{numericalaccuracy}
First, we assess the predictive capability of the failure function from a numerical accuracy point of view. For this purpose, we compute the stability-limiting critical strains ---for biaxial deformations referred to sets of axes that are not included in the sampling set --- and compare with directly-calculated values from a phonons-based stability analysis.  As shown in Fig.~(\ref{Validate}), the agreement between the predicted stability limits from the failure model and the calculated values is very close. This, in particular,  highlights the efficacy of the symmetry-invariants based representation in constitutive modeling, since the representation of the failure function utilized the explicit calculations \textit{only} along the two symmetry-directions of graphene. Previously also, in hyperelastic constitutive modeling of graphene \cite{kumar2014on}, we showed that the representation in terms of symmetry-invariants not only reduces the number of constants in the model; but it also elucidates the underlying hyperelastic softening behavior of the graphene lattice.\\ \\
\begin{minipage}{\linewidth}
\centering  \includegraphics[scale=0.6]{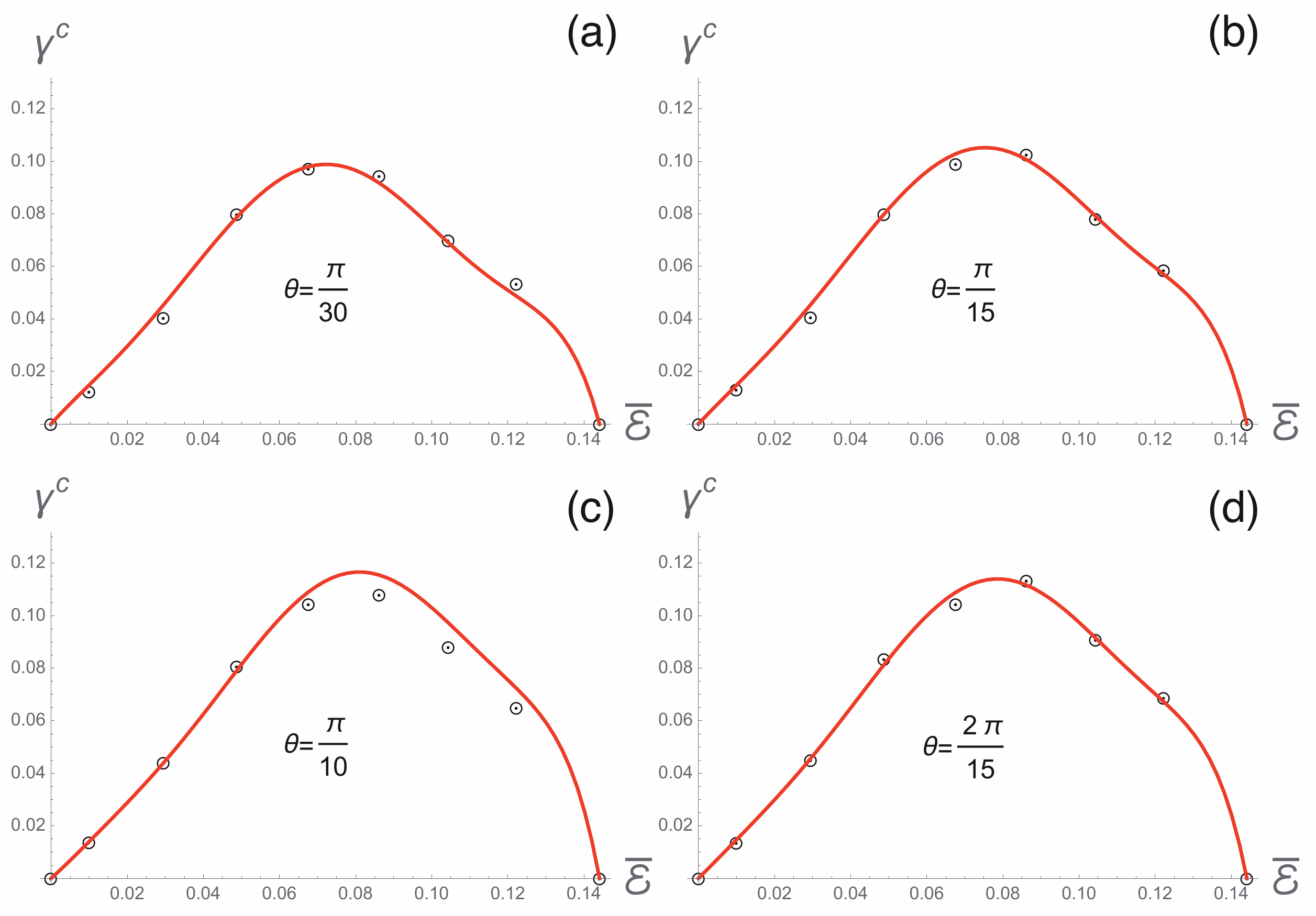}
\captionof{figure}{\small Shown are the stability limits for deviatoric stretch angles (a) $\theta= \pi/30$ (b) $\theta= \pi/15$ (c) $\theta= \pi/10$ and (d) $\theta= 2\pi/15$.  These directions were not included in the interpolation procedure. The interpolated stability limits (shown by solid red lines) by the model are in close agreement with the directly-calculated values (shown by black dotted circles) obtained by lattice-dynamical stability analysis.}
\label{Validate}
\end{minipage}
%
\subsection{Comparison with acoustic tensor ($\Lambda$-) analysis --- elastic (material) instabilities in tension}
\label{acoustictensor}
In a previous work \cite{kumar2014on}, we showed that both the 2D area modulus $\kappa(\epsilon_a) = \left[\frac{\partial \text{tr} \pmb \sigma/2}{\partial \epsilon_a} \right]_{\ln \lambda =0} $ and the 2D shear modulus $\mu(\epsilon_a)$ of graphene experience a progressive softening with increasing areal strain. The progressive softening of the elastic moduli is often responsible for long-wavelength material instabilities of the underlying lattice. The acoustic tensor derived from the constitutive response can be used to determine the macroscopic instabilities of the graphene lattice, arising from the hyperelastic softening. If, for some pair of unit vectors $\mathbf m$ and $\mathbf n$, 
\begin{equation} \Lambda (\mathbf m, \mathbf n) =(\mathbf m \otimes \mathbf n) : \mathbb A : (\mathbf m \otimes \mathbf n) \le 0,
\end{equation} then the material is elastically unstable; here  $\mathbb A = \partial^2 \psi/ \partial \mathbf F^2$ is the acoustic tensor (see \cite{kumar2014on} for detailed evaluation of $\mathbb A$).   Employing the acoustic tensor-based analysis, we determine the stability-limiting critical strains for varying biaxiality in a biaxial deformation with stretch axes $\mathbf r_1 - \mathbf r_2$ aligned along zigzag-armchair ($\mathbf e_1 -\mathbf e_2$) set of axes, i.e., 
\begin{equation}
\mathbf E^{(0)} = \mathcal E_{11}^{0} \mathbf e_1 \otimes \mathbf e_1 +\mathcal E_{22}^{0} \mathbf e_2 \otimes \mathbf e_2.
\end{equation}
We plot the resulting limiting values of $\mathcal E_{11}^{0},\ \mathcal E_{22}^{0}$ ---obtained from acoustic tensor analysis --- and superpose them on the  phonon-instability contour as shown in Fig.~(\ref{superpose}).  The agreement is close, except in the neighborhood of the equi-biaxial deformation. 
For near equi-biaxial deformations($\mathcal E_{11}^{0} \approx \mathcal E_{22}^{0}$),  the instability of the graphene lattice is not of elastic nature but corresponds to microscopic soft-phonon modes (see Yevick \& Marianetti \cite{marianetti2010failure}), which can not be detected via acoustic-tensor analysis. Following a simple algebraic procedure, it can also be shown that in the long wavelength limit, the phonon-based criterion and the acoustic tensor based condition become equivalent, and therefore in the case of elastic instabilities, the agreement between results of the two analyses is close.
\subsection{Comparison with buckling analysis --- structural instabilities in compression}
\label{buckling}
The 2D acoustic tensor--- based on the in-plane constitutive response--- captures material instabilities of macroscopic (elastic) nature only.  However, in addition to material instabilities, structural instabilities of non-material nature may also arise. These instabilities are the buckling instabilities ensued by compressive stresses (precisely speaking, when at least one of the eigenvalues of $\pmb \sigma$ becomes less than zero). The buckling modes on the instability contour are identified as the state of strains at which one of the principal stress vanishes. For example, in biaxial deformations with stretch axes $\mathbf r_1 - \mathbf r_2$ aligned along zigzag-armchair set of axes, the buckling stability-limits are given by strain trajectories (shown in Blue in Fig.~(\ref{superpose})) calculated from the constitutive model for uniaxial tensile stress along these two special directions. 

It is interesting to see that, in the small strain limit, the two buckling stability-limiting contours are related to Poisson contraction along the zigzag and armchair directions. Taking advantage of isotropy at small strains,  linearization of the constitutive relation in terms of principal stresses and strains gives
\begin{equation}
\mathcal S_{11}^0 = E_{2D}(\mathcal E_{11}^0+ \nu_{2D} \mathcal E_{22}^0);
\end{equation}
and 
\begin{equation}
\mathcal S_{22}^0 = E_{2D} (\mathcal E_{22}^0 + \nu_{2D} \mathcal E_{11}^0).
\end{equation}
Here the asymptotic value of the in-plane Poisson ratio emerging from our constitutive model is $\nu_{2D} = 0.205$ \cite{kumar2014on}. The asymptotic principal strain trajectories corresponding to the condition of buckling instability are obtained as follows.\\ \\
\textbf{Buckling in armchair direction:}
\begin{equation}
\mathcal E_{22}^0 = -\nu_{2D}\ \mathcal E_{11}^0,\, \mathcal E_{11}^0 > \mathcal E_{22}^0.
\label{nu1}
\end{equation}
Thus, we obtain the following relation between $\gamma$ and $\bar{\mathcal E}$:
\begin{equation}
\gamma = \frac{1}{2} (\mathcal E_{11}^0  - \mathcal E_{22}^0 ) = \frac{(1+\nu_{2D})}{(1-\nu_{2D})} \bar{\mathcal E}.
\label{buckle}
\end{equation}
\textbf{Buckling in armchair direction:}
\begin{equation}
\mathcal E_{11}^0 = -\nu_{2D}\ \mathcal E_{22}^0,\, \mathcal E_{22}^0> \mathcal E_{11}^0.
\label{nu2}
\end{equation}
and again, we get the same relation between $\gamma$ and $\bar{\mathcal E}$ as in equation~(\ref{buckle}) for buckling instability. The buckling stability-limiting trajectory on the $\gamma- \bar{\mathcal E}$ plane in the asymptotic limit corresponds to a straight line emanating from origin with slope $(1+\nu_{2D})/(1-\nu_{2D})$, and as noted previously $E_{2D}$ is the elastic modulus of graphene in uniaxial strain.\\
%
\begin{minipage}{\linewidth}
\centering
\includegraphics[scale=0.5]{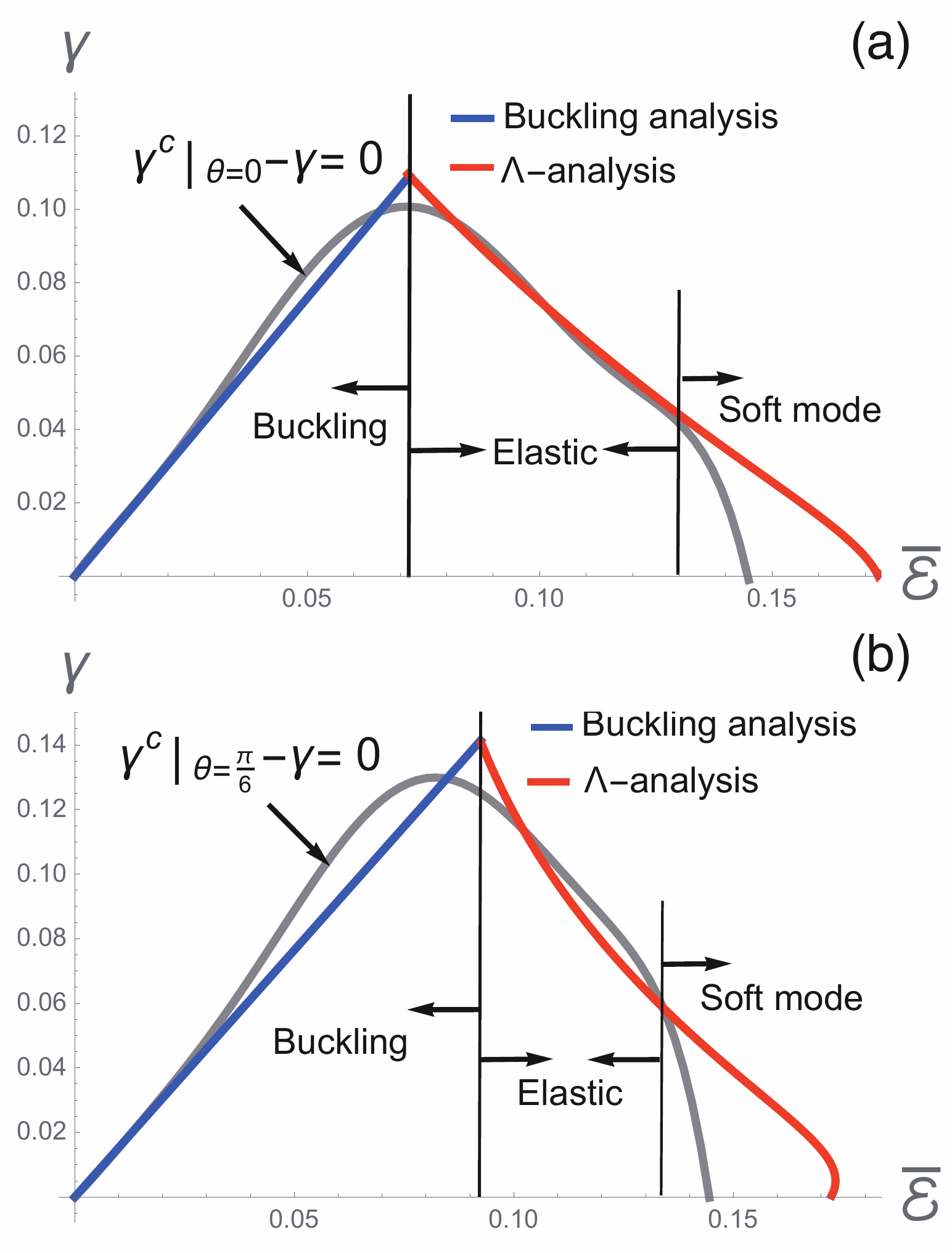} 
\captionof{figure}{\small (a) The 2D radial section of the limit surface corresponding to $\theta=0$ (zigzag direction), i.e., $\gamma = \gamma^c\rvert_{\theta=0}$, superposed onto the elastic-stability limit curve obtained from $\Lambda-$ analysis and the buckling-instability curve obtained from Poisson analysis. (b) The 2D radial section of the limit surface corresponding to $\theta=\pi/6$ (armchair direction), i.e.,$\gamma = \gamma^c\rvert_{\theta=\pi/6}$, superposed onto the elastic-stability limit curve. Intersection of Poisson analysis and the acoustic instability analysis marks the uniaxial tension failure. Note that in large, \textit{nearly} equi-biaxial deformations, the phonon-instability precedes the elastic instability (conforming with Yevick \& Marianetti's \cite{marianetti2010failure} calculations), indicating that in such deformations, lattice instability is due to a short-wavelength instability.}
\label{superpose}
\end{minipage} \\ \\
\section{Implementation of the limit surface in a continuum FEA scheme}
\label{FEA}
\subsection{Outline of the methodology}
\label{outline}
In the following, we outline a systematic procedure for implementation of the proposed limit function in a finite element analysis in order to provide an on-the-fly assessment of incipient material failure during monotonic loading.
\begin{itemize}
\item First, using the continuum-level FEA based on the nonlinear constitutive model of equation~(\ref{wcstress}) and equation~(\ref{cauchy}) ---  we obtain the macroscopic deformation field associated with the given boundary-value problem at the current load/displacement increment. Note that the stress/strain field in general is inhomogeneous; however, if the spatial variation of elastic fields varies sufficiently slowly, each material element in a  sufficiently fine mesh can be fairly assumed to be in the state of homogeneous deformation, with the local stress/strain determined from the macroscopic deformation field.

\item  Based on the local stress/strain value, we make on-the-fly assessment of the failure function at the material elements in the mesh for the current load/displacement step. For each increment, the material elements with the minimum values of the failure function are identified as the `weakened-spot'.
\item The value of the critical macroscopic load/displacement is determined by identifying the load/displacement increment along the loading path at which the failure function at the weakened-spot first vanishes. The weakened-spot is then declared as a `failed region', and the corresponding load/displacement can be considered as failure load/displacement.
\item We determine value of the critical stress/strain by locating the point of intersection of the stress/strain trajectory traced by the `failed' material elements during the course of loading with the failure surface.
\item  Finally, we check if the failure surface precedes or coincides with the acoustic-instability surface at the point of intersection. This allows us to determine whether the instability is a soft mode instability or an elastic instability, respectively.
\end{itemize}
\textbf{Remark 1 :} Since the  phonons are defined with respect to an ideal infinite crystal without any surface or boundary, the application of phonons to stability detection in real boundary-value problems requires a little care. For this we need an volume/area element, called a RVE, the size of which should be sufficiently large so that
(1) it accommodates the unstable mode (2) the mechanical response is described well by the continuum constitutive relation. As long as in a given boundary value problem, we can identify such an RVE, the phonons can be used  an indicator of material stability within the RVE. Once such an RVE is identified, the instability can be detected by phonon-based lattice instability analysis.\\ \\
\textbf{Remark 2 :} Since a soft mode phonon instability result from dynamical growth of certain modes of lattice-vibrations, a constrained boundary may suppress triggering of a short-wave instability in a nearby material element even if the local state of strain in the material element has reached the limiting condition. Therefore, the use of the limit surface in identification of an incipient failure is rigorously valid only in the interior, sufficiently far away from the boundary conditions.\\ \\
\textbf{Remark 3 :} An FEA simulation based on the continuum constitutive model of Sec.~(\ref{straintostressspace}) collapses immediately after an elastic instability is encountered, since the hyperbolicity of the equations of motion no longer remains valid. However, since a continuum model is blind to short-wave instabilities, an FEA simulation continues even if a short-wave instability is encountered, until it reaches the point of an elastic instability (see Fig.~(\ref{Pvsd}) in the following section).
\subsection{Prediction of incipient failure in blistering of a defect-free graphene sheet}
\label{example}
Employing the proposed scheme, we analyze the incipient failure of a defect-free graphene sheet in  FEA simulations of idealized bulge-type tests. A bulge test involves blistering of a graphene sheet clamped at its boundary on top of a microcavity containing a pressurized gas. Two different geometries for the constrained boundary are considered in our work: elliptical and circular, while the characteristic dimension $\zeta = \sqrt{a b}$ remains same, where $a$ and $b$ are the sheet dimensions along the zigzag and armchair directions, respectively. For elliptical geometry, we choose $a=1.0\,\mu\text{m}$ and $b=0.5\, \mu\text{m}$ for the first configuration, and $a=0.5\,\mu\text{m}$ and $b=1.0\, \mu\text{m}$ for the second configuration; and for circular geometry, we choose $a=b=1/\sqrt{2}\, \mu\text{m}$. \\ \\ 
\begin{minipage}{\linewidth}
\centering
\includegraphics[scale=0.6]{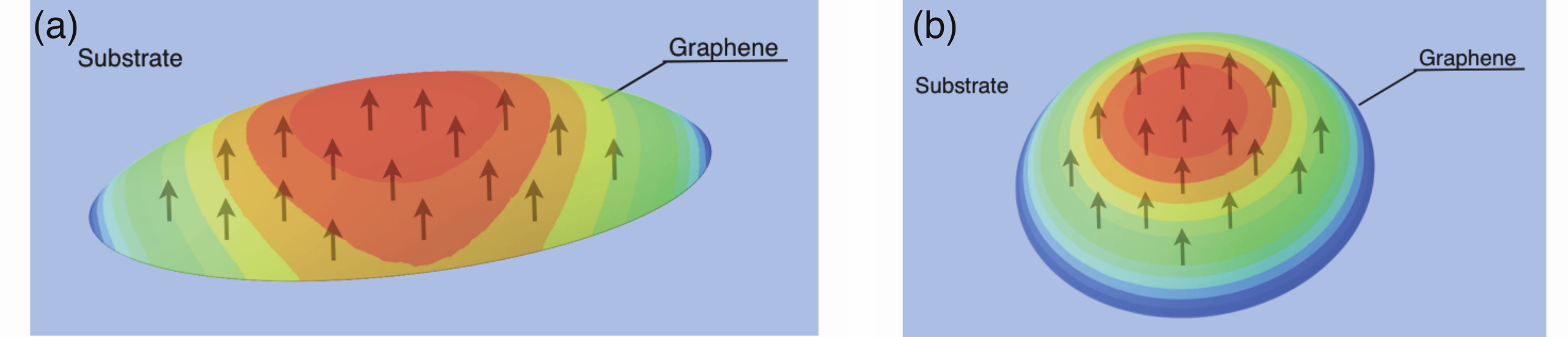} 
\captionof{figure}{\small Schematic of blistering (a) of an elliptic membrane and (b) of a circular membrane.  The membrane is clamped at the boundary, and is subjected to a pressure (denoted by arrows) from below, resulting in the bulged configuration as shown in the figure. The state of strain at the center of the membrane is equi-biaxial in case of a circular membrane, whereas in case of elliptic membranes, it is biaxial, with larger strain in the minor direction.}
\label{blister_schematic}
\end{minipage}
\\ \\
Our objective is to determine the macroscopic and microscopic conditions at which the graphene sheet fails. Numerically implementing the constitutive response and the failure function in a commercial FEA program $\scriptstyle{ABAQUS}$, we explicitly determine the critical pressure-difference $\Delta P_0^{\mathrm c}$ and the critical central deflection $\delta_c^{\mathrm c}$; the location of unstable material element; the local stress and strain at the unstable element; and the nature of instability. We demonstrate this procedure systematically for the blistering of the circular and elliptical graphene sheets.

\subsection*{(a)  Location of unstable material element and geometry of failed region}
To locate the unstable material element, we monitor the evolution of the limit function $\mathcal F = \gamma^c(\mathcal E, \Theta(\theta)) - \gamma$ over the suspended graphene sheet, as a function of externally-applied pressure-difference (shown in Fig.~(\ref{blister_fea})). The material element at which the limit function first vanishes denotes the unstable region, which will subsequently fail. This unstable region determines the geometry of the failed region. For example, in blistering of a circular graphene sheet --- based on the negativity of the unstable region --- we would expect that the fracture initiation occurs at the center in the form of a nearly circular void.\\ \\
\begin{minipage}{\linewidth}
\centering
\includegraphics[scale=0.55]{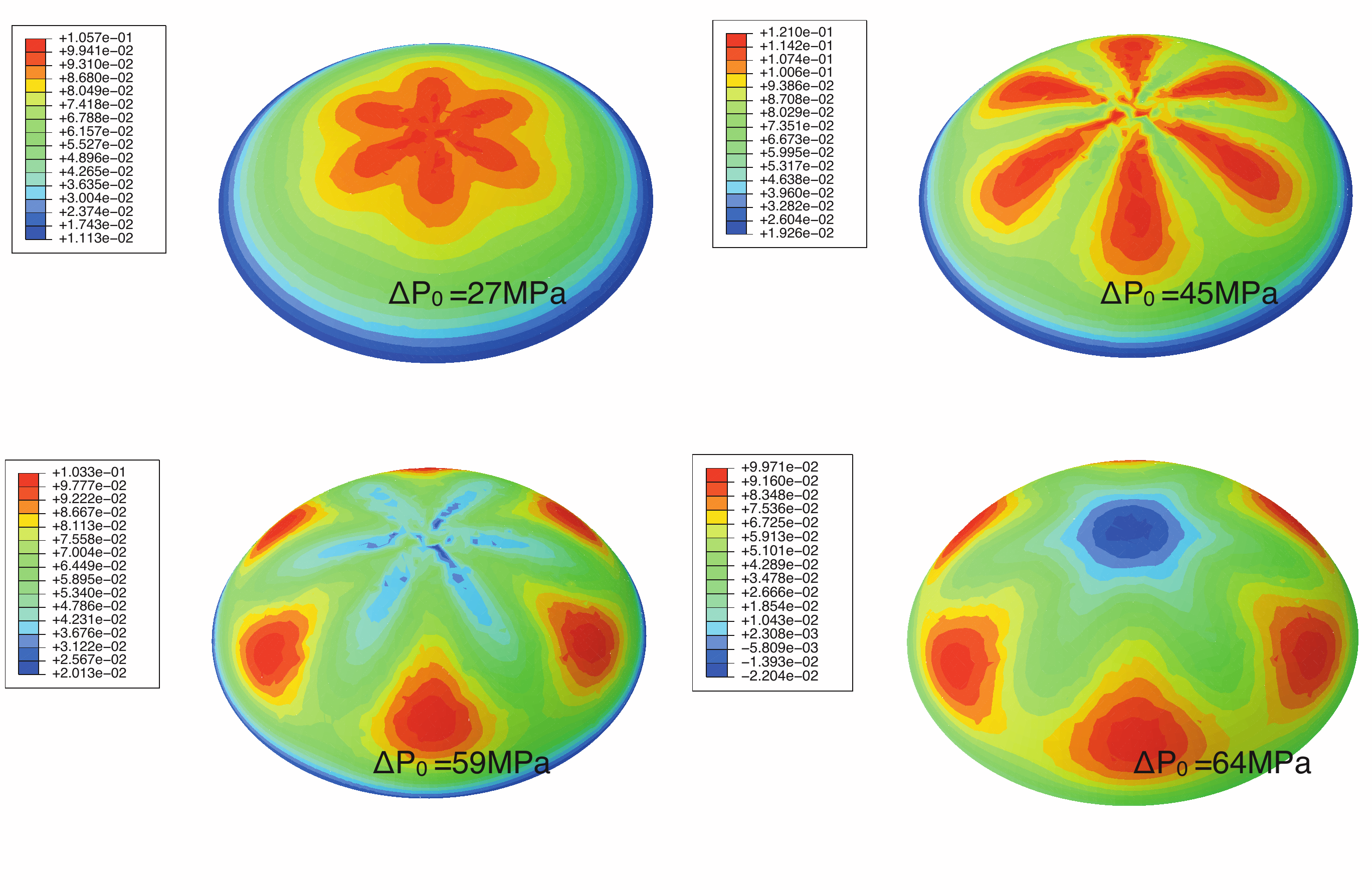} 
\captionof{figure}{Sequence of images showing the evolution of the limit function $\mathcal F = \gamma^c((\mathcal E, \theta)) - \gamma$ with pressure-difference $\Delta P_0$ across the circular membrane. The central region with negative value of the limit function indicates the material elements that have reached the lattice-stability limit, which are at the verge of mechanical failure.}
\label{blister_fea}
\end{minipage}
\subsection*{(b) Pressure-difference and central deflection at the onset of instability}
The critical pressure-difference ($\Delta P_0^{c}$) and the critical central deflection ($\delta_c^{c}$) are determined by locating the onset of instability on the pressure-deflection response curve (shown in Fig.~(\ref{Pvsd})). It should be noticed that the hyperbolicity of elastic-wave equations is essential for a finite element analysis step to continue; if an elastic instability is  first reached along a loading path, as in the case of elliptical graphene sheet, then the condition of hyperbolicity ceases to hold, and the FEA step immediately terminates. On the other hand, if a soft-mode instability is first reached on the loading path, the elastic-wave equation remains hyperbolic, and the FE steps continues until an elastic-instability along the loading path is reached. For example, during the blistering of a circular sheet, the microscopic soft-mode instability takes place prior to the elastic instability; the simulation continues past the soft-mode instability, terminating at the point of elastic-instability. We have indicated the initial encounter with both the soft-mode and the elastic instability on the pressure-deflection response curve. From Fig.~(\ref{Pvsd}), it clearly emerges that in the absence of the limit surface, an FEA simulation is bound to overestimate the mechanical strength of a crystal when the strength-limiting mechanism along the loading path is a short-wave instability.\\ \\
\begin{minipage}{\linewidth}
\centering
\includegraphics[scale=0.35]{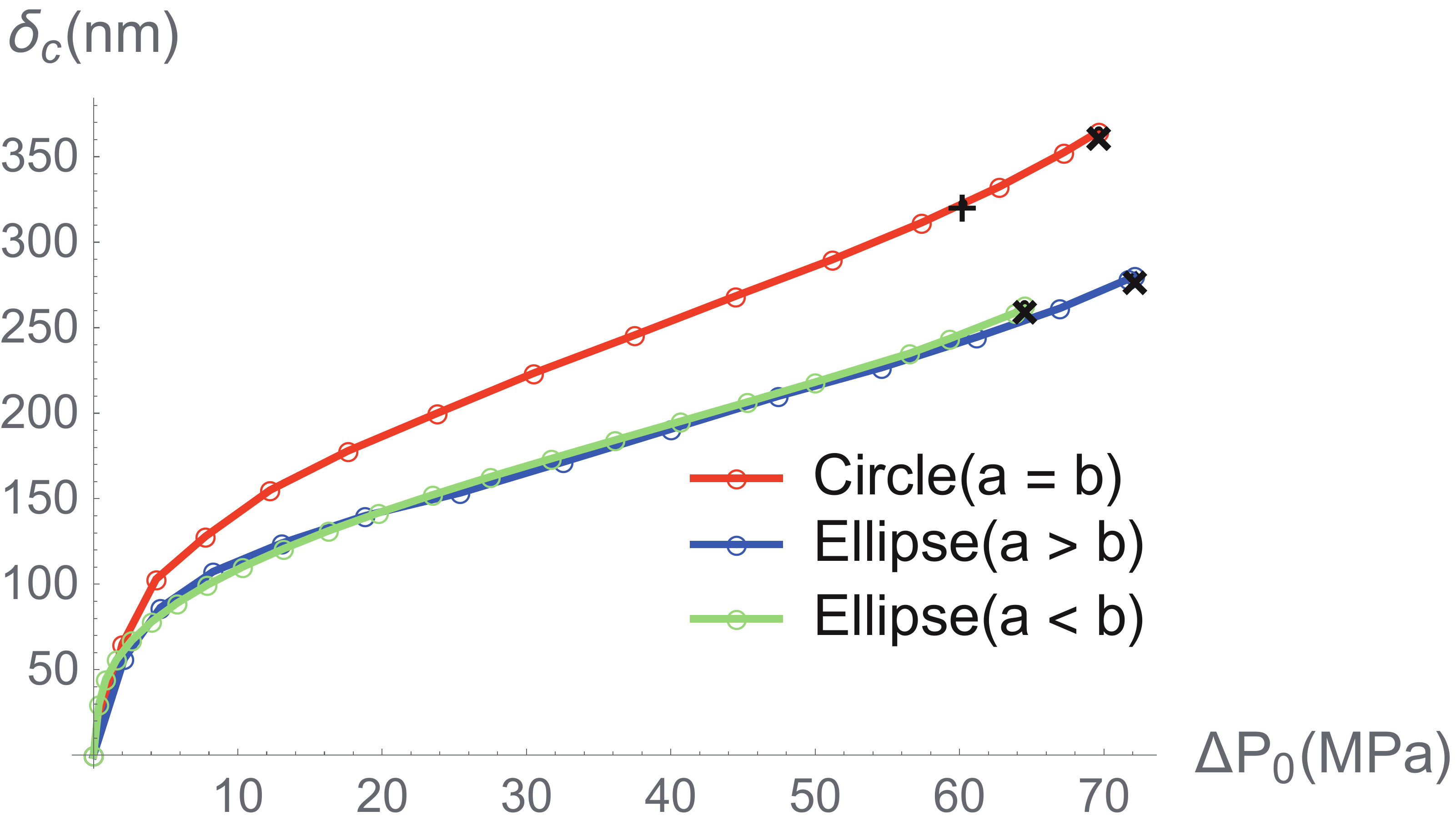} 
\captionof{figure}{Pressure vs deflection response curves obtained from the elliptical and the circular blister simulations. On these curves,  the long-wave (elastic) instabilities are located by a $\mathbf{\times}$; while the location of the short-wave (soft-mode) instability is denoted by $\mathbf{+}$. }
\label{Pvsd}
\end{minipage}
\subsection*{(c) Local stress/strain at the unstable material element}
The local principal strains at the critical material element immediately after the onset of instability, $\mathcal E_{\text{max}}^c$ and $\mathcal E_{\text{min}}^c$ ,  are obtained by tracing the trajectory in the strain space, and obtaining the intersection of this trajectory with the limit surface, as shown in Fig.~(\ref{crit_ellipse_traject}). Similarly, the local principal stresses at the critical element at the onset of instability, $\mathcal S_{\text{min}}^c$ and $\mathcal S_{\text{min}}^c$, are obtained by tracing the trajectory and locating its intersection with the limit surface in the stress space. These values are tabulated in Tab.~(\ref{stressstrain}).
\begin{table}[htbp!]
\begin{center}
\begin{tabular}{c c c c c c c c c c }
\toprule[1.5pt]
Geometry of constrained boundary& $\mathcal S_\text{max}^c$(N/m) & $\mathcal E_{\text{max}}^c$ & Nature of instability & $\Delta P_0^{c}$  (MPa)\\ \\
\hline 
\hline
 & & & & & & & \\
Elliptical $(a > b)$ & 37.74 & 0.20 & Elastic & 72 \\
Elliptical $(a < b)$ & 34.31 & 0.175  & Elastic & 64 \\
Circular  $(a = b)$ & 31.28 & 0.1475 & Soft-mode & 62 \\
\bottomrule[1.65pt]
\end{tabular}
\end{center}
\caption{ Local major principal stress and local major principal strain at the unstable material element as calculated by finite element analysis. It is noted that the strength is determined by a complex interplay between anisotropy, the geometry of the boundary condition and the nature of instability governing the failure.}
\label{stressstrain}
\end{table}

From Tab.~(\ref{stressstrain}), we note that the limiting stress and strain in the blistering simulations are inherently dependent on the geometry of the constrained boundary. Due to the fact that for certain geometries, e.g., when $a \approx b$, the failure is mediated by a microscopic soft-mode instability, the sensitivity of the failure stress/strain on the geometry of the constrained boundary can not be explained by the material anisotropy or the nonlinear softening in the material response alone. Thus, in a global sense, the proposed failure model is a useful method for assessment of correlation between the failure stress/strain and extrinsic factors. 

\subsection*{(d) Nature of instability:  long-wave versus short-wave instabilities}
The proposed limit function captures lattice instabilities of all kinds; however, without resolving whether it is  macroscopic (elastic) or microscopic (soft mode) in nature. By combining the proposed limit function with the acoustic tensor analysis, the distinction between macroscopic and microscopic instabilities can be readily made. To this end, we use the fact that an acoustic-tensor analysis captures macroscopic instabilities but not the microscopic ones \cite{ziman1972principles}. Thus, if the instability is indeed a macroscopic instability, the predicted onsets of instability from the failure function and the acoustic-tensor analysis coincide; however, when the instability is a soft mode, the predicted onset of failure from the limit-function precedes that from the acoustic-tensor analysis (see Fig.(\ref{crit_ellipse_traject})). \\ \\
\begin{minipage}{\linewidth}
\centering
\includegraphics[scale=0.55]{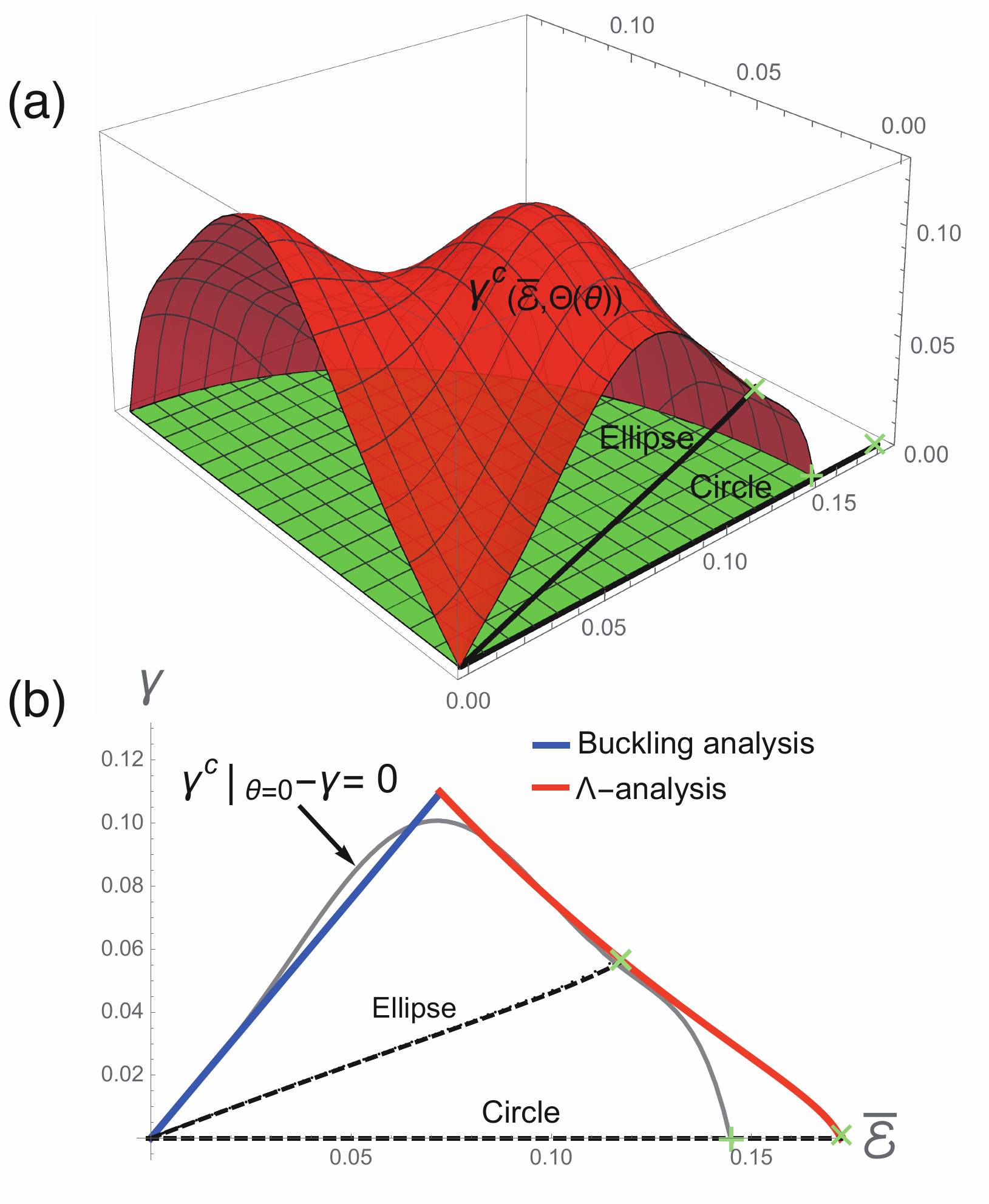} 
\captionof{figure}{ (a) Shown are the deformation paths followed by the critical material elements during the elliptical and the circular blister simulations, along with the lattice-stability limit surface. The intersection of the deformation path with the limit surface indicates the onset of failure. (b) Deformation paths for the elliptical ($a < b$) and the circular bulges, superposed on the $\theta =0$ radial section of the limit surface.  The long-wave (elastic) instabilities are located by a $\times$; while the location of the short-wave (soft-mode) instability is denoted by $+$.  For the elliptical membrane, the points of intersection of the trajectory with the limit surface and the elastic instability curve coincide, indicating that the failure is of elastic nature; while for the circular membrane, the intersection with the limit surface precedes that with the elastic instability curve, indicating that the failure is triggered by a soft mode instability.}
\label{crit_ellipse_traject}
\end{minipage}
\\
\section{Conclusion and discussion}
\label{conc}
Onset of inelasticity or `failure' in  defect-free graphene crystals ensues from the deformation-induced instabilities of the underlying lattice. Though lattice-instabilities of all kinds can be assessed via an elaborate atomistic-level lattice-dynamical simulation,  there remains a lack of a general continuum criterion capable of describing the conditions for the onset of inelasticity along an arbitrary deformation path, parametrized in terms of stress or strain.  In addition to non-linearity and anisotropy in material response at large-deformations, a major difficulty in this regard arises due to the fact that different lattice-instability mechanisms govern the failure in along different loading paths.\\
For a majority of deformation paths, material failure in crystals arises from a long-wave or elastic instability. Such elastic instabilities can easily be predicted by a continuum acoustic tensor analysis once an appropriate hyperelastic constitutive relation for the mechanical response of the material has been formulated \cite{hill1962acceleration, hill1977principles}. Further, an elastic instability is readily seen in an FEA simulation, since upon reaching an elastic instability, the hyperbolicity of equations of motion ceases to hold and the unstable material element undergoes excessive distortion, annihilating the load carrying capacity of the material. However, there also remains a small but finite domain of deformation paths along which a short-wave (also called soft mode) emerges before an elastic instability is reached, and thus limiting the strength along those deformation paths \cite{marianetti2010failure}. Unfortunately, there exists no equivalent continuum criterion that could describe the onset of short-wave instabilities, and therefore a continuum FEA simulation remains blind to such instabilities. Thus, in the absence of a multi-scale resolution, an FEA simulation is bound to overestimate the mechanical strength of a crystal when the strength-limiting mechanism along the loading path is a short-wave instability. \\
Here, we have presented for the first time the notion of a comprehensive lattice-stability limit surface --- a continuum parametrization which describes the loss of internal stability of a defect-free graphene sheet --- both in terms of stress and strain --- with respect to perturbations of all kinds. Further, on this limit surface, we have identified the distinct regions that correspond to fundamentally different kinds of lattice-instability mechanisms. Such a continuum parametrization is valuable for several reasons:
\begin{itemize}
\item First, it duly accounts for all the complexities in crystal response, such as non-linearity, anisotropy and multiplicity of instability mechanisms, via one general continuum criterion: an instability appears when the magnitude of deviatoric strain reaches a critical value that depends upon the mean hydrostatic strain and the directionality of stretch; and the nature of this instability depends upon the mean hydrostatic strain level.
\item Secondly, despite being based on a continuum formulation, it possesses a multi-scale resolution. It can address the incipient failure of defect-free graphene crystals under an arbitrary loading condition and at realistic length scales --- by the virtue of its continuum description --- while the built-in microscopic lattice dynamical theory of stability analysis allows us to capture instabilities as-short-as the size of few unit-cells.
\item It is based on a novel interpolation scheme based on the symmetry-invariants of the logarithmic strain measure, which reduces the sampled deformation paths required for representation to a bare minimum of two: one along the zigzag direction and the other along the armchair direction.
\item Owing to its continuum representation, the limiting criterion is easily implemented in a continuum FEA simulation. Such an implementation is capable of performing an on-the-fly lattice stability analysis with an on-going finite element calculation, and can predict the location of the unstable material point, the load/displacement at the onset of instability, the local stress/strain, and the character of the instability, i.e., a short-wave or a long-wave in a realistic boundary-value problem.
\end{itemize}
Employing the limit surface in an FEA simulation, we analyzed the failure of a defect-free suspended graphene sheet in the finite elements simulation of idealized bulge-type tests. Our simulations considered two different geometries of the constrained boundary, elliptical and circular, but with the same suspended area. We showed how different types of local lattice instabilities are reached in the respective boundary value problems under different levels of differential pressure and overall deflection. From these examples, it emerged that, in the absence of a multi-scale resolution, a continuum-level finite element analysis is bound to overestimate the strength of graphene along certain deformation paths. Further, we showed that the proposed limit surface, owing to its continuum representation and multi-scale resolution, adequately remedies this deficiency.\\
In a recent work \cite{kumar2015strain}, we have demonstrated the utility of the proposed limit surface in identifying a strain-shielding effect initiated by mechanically-activated covalent interactions at the graphene-indenter interface during nano-indentation of graphene. Employing the limit surface as the failure criterion in a FEA simulation, we explained how this strain-shielding effect can lead to a delay in the onset of instability in graphene during nano-indentation.

\end{document}